\documentclass[pre,nofootinbib,floatfix, superscriptaddress,a4paper]{revtex4}

\usepackage{graphicx,setspace}
\usepackage{amsfonts,amsmath,amssymb,bm}
\usepackage{standalone}
\usepackage{setspace}
\usepackage{natbib}
\usepackage[dvips,usenames]{color}

\newcommand{\br}{{\bf r}}
\newcommand{\bx}{{\bf x}}
\newcommand{\be}{{\bf e}}

\begin{document}

\title{Self-diffusiophoresis induced by fluid interfaces}

\author{P. Malgaretti}
\email[corresponding author: ]{malgaretti@is.mpg.de}
\author{M.N. Popescu}
\author{S. Dietrich}
\affiliation{Max-Planck-Institut f\"{u}r Intelligente Systeme, Heisenbergstr. 3, D-70569
Stuttgart, Germany}
\affiliation{IV. Institut f\"ur Theoretische Physik, Universit\"{a}t Stuttgart,
Pfaffenwaldring 57, D-70569 Stuttgart, Germany}

\begin{abstract}
 The influence of a fluid-fluid interface on self-phoresis of  chemically active, 
axially symmetric, spherical colloids is analyzed. 
Distinct from the studies of self-phoresis for colloids trapped at fluid interfaces or in the vicinity of hard walls, here we focus on the issue of self-phoresis \textit{close} to a fluid-fluid interface.
In order to provide physically intuitive results highlighting the role played by the 
interface, the analysis is carried out for the case that the symmetry axis of the 
colloid is normal to the interface; moreover, thermal fluctuations are not taken into 
account. 
Similarly to what has been observed near hard walls, we find that such colloids can be set into motion even if their whole surface is homogeneously active. This is due to the anisotropy along the direction normal to the interface owing to the partitioning by diffusion, among the coexisting fluid phases, of the product of the 
chemical reaction taking place at the colloid surface. Different from 
results corresponding to hard walls, in the case of a fluid interface the 
direction of motion, i.e., \textit{towards} the interface or \textit{away} from it, can 
be controlled by tuning the physical properties of one of the two fluid phases.
This effect is analyzed qualitatively and quantitatively, both by resorting 
to a far-field approximation and via an exact, analytical calculation which 
provides the means for a critical assessment of the approximate analysis.
\end{abstract}

\maketitle

\section{\label{intro} Introduction}

The active motion of micro- and nano-sized particles has gained significant interest 
(see recent reviews such as Refs. \cite{LaugaRev,ebbens,Gompper2015_rev,Bechinger_RMP2016}) due to a wide, associated spectrum of 
applications including drug-delivery 
systems~\cite{Wang_2017a,Wang_2017b,Wang_2017c} as well as due to the ability to mimic motile biological cells such as 
bacteria~\cite{Poon}. In view of these perspectives diverse approaches aiming at the 
understanding of active motion of micro- and nano-sized particles have been put 
forward; a detailed account of them can be found in Ref.~\cite{ebbens}. A 
particularly promising approach is the case of 
self-phoretic particles~\cite{Paxton2004,Golestanian2005,Howse2007,Kapral2007}, which 
attain net motion by means of chemical reactions catalyzed on parts of  their 
surface.
Indeed, if the rate of the chemical reaction on the surface of such particles is 
distributed 
inhomogeneously, local density gradients in the concentration of reactants and products 
lead to imbalances in the local chemical potential and pressure which eventually, by 
generating fluid velocity profiles, set the particle into motion. Several 
theoretical~\cite{Golestanian2005,Kapral2007,
Julicher,Popescu2009,Popescu2010,Lowen2011,Seifert2012a,Koplik2013,Kapral2013,Brown2017}
as well as 
experimental~\cite{Paxton2004,SenRev,ebbens,Howse2007,Golestanian2012,Fisher2014,
Stocco} studies have 
characterized the performance of such systems. Recently it has been shown 
that the dynamics of Janus particles (i.e., catalytic particles with the catalyst 
distributed only across a part of their surface) is quite sensitive to the presence of 
boundaries, obstacles, or other means which can distort the density profiles of the reaction 
products and 
the local velocity profiles, hence inducing a modulation of the net displacement of such 
kind of particles. Resembling the well-documented ``wall-attraction'' for 
micro-organisms and mechanical swimmers (see, e.g., Refs.~\cite{Lauga2006,Lauga2008,Lauga2014b,Gompper2015,chinappi2016}), for self-phoretic particles it has been shown that, in addition to the modulation of the velocity~\cite{Popescu2009,Yariv2016,Chang2016}, wall-bounded steady states emerge from the interplay of hydrodynamic and phoretic interactions with the wall (i.e., via the wall-induced distortions of 
the hydrodynamic flow and of the distribution of the chemical species)~\cite{Uspal2015,Yariv2016,Ibrahim2016,Sharifi-Mood2016}. This can be exploited to achieve a guided motion of the active colloid 
\cite{Howse2015,Simmchen2016,Uspal2016}. The emergence of similar effective interactions 
between pairs of active colloids have been studied theoretically in Refs.~\cite{Popescu2010,Michelin2015,Koplik2016b,Parvin2016}. Recently, experimental~\cite{Stocco,Isa2017} and theoretical~\cite{Masoud2014,Wurger2014,Stark2014,Malgaretti2016,Dominguez2016,Dominguez2016_2} 
studies have started to tackle the issue of motion of active colloids near or trapped at 
a liquid-fluid interface. With respect to the single particle motion, a certain increase 
in the persistence length of the self-phoretic motion has been 
reported for particles trapped at the interface~\cite{Stocco,Isa2017,Malgaretti2016}. Instabilities of interfaces covered by 
active particle, which can be considered themselves as surfactants, have been reported in 
Ref.~\cite{Stark2014}. Alternatively, the motion of active particles which induce 
Marangoni stresses at the interface via the chemical species or heat they release (or absorb) has 
been studied both in terms of the emergence of motion for a single 
particle~\cite{Masoud2014,Wurger2014,Dominguez2016} as well as concerning the issue of 
the dynamics and stability of monolayers of such particles trapped at the interface 
\cite{Masoud2014,Dominguez2016_2}. (See also earlier studies on thermocapillary motion near interfaces, 
e.g., in Ref.~\cite{Leshansky1997}.)

In this context, here we address self-phoresis of a catalytic particle in the vicinity 
of a 
fluid-fluid interface. First, in order to emphasize the influence of the interface,  
we study the case of a particle which is homogeneously catalytic. 
In order to keep the system as simple as possible, in the following we assume that 
the concentration of reactants is kept constant in space and time.
In such a situation the particle releases the products of the catalytic reaction isotropically. 
Therefore, even though the system is kept out of 
equilibrium, due to the isotropic coverage of the catalyst in the bulk there is no symmetry 
breaking\footnote{Here we assume that the 
advection of the products and reactants by the hydrodynamic flow is negligible 
compared with their diffusion so that motion of the particle due to a spontaneous symmetry breaking 
\cite{Michelin2013,Buyl2013} does not occur.} and hence no net displacement. However, the 
presence of an interface breaks the homogeneity of the transport coefficients (i.e., 
the diffusivities of the reaction products in the two fluid phases). This leads to an 
inhomogeneous distribution of the concentration of the reaction products along the 
interface normal.  Accordingly, such an inhomogeneous density profile 
leads to an interface-induced phoresis, the direction of which is normal to the interface, 
similarly to what has already been reported for the case of a hard 
wall~\cite{Uspal2015,Yariv2016}. Interestingly, in the present case the sign 
of the resulting velocity depends not only on the surface properties of the particle, as 
it is the case near a hard wall, but it depends on both the contrast between the diffusivities and the distinct solvability of the 
catalysis products in the two fluid phases. 

If the catalyst is not homogeneously distributed along the surface of the particle, a net 
motion will arise due to self-diffusiophoresis even in a homogeneous fluid. Accordingly, 
if such particles are close to a fluid-fluid interface, a competition arises between 
the intrinsic motility and the one induced by the interface. Since the motility induced by 
the interface is directed solely along its normal, in order to grasp the interplay between 
the self-phoresis and the interface-induced phoresis we focus on the case of 
an asymmetrically coated colloid (ideal Janus particle) the axis of symmetry of which is 
parallel to the normal of the interface.

The paper is organized as follows.
In Sec.~\ref{model} we formulate the model describing the dynamics of active 
colloids close to a fluid interface. In Sec.~\ref{Results} we study the velocity of 
these active colloids by using an approximate far-field expansion as well as 
an exact solution. While the approximate far-field expansion allow us to 
straightforwardly  grasp the phenomenology emerging from 
the dynamics of active colloids close to fluid interfaces, the exact 
solution, by providing quantitatively reliable results, allow us to critically asses 
the strengths, as well as the shortcomings, 
of the approximate far-field approach. Finally in Sec.~\ref{sum} we provide 
concluding remarks. The details of the necessary calculations beyond the ones presented in the 
main text are included in the Appendices~\ref{app:bi--polar}-\ref{app-slip}.

\section{\label{model} Model system}
\begin{figure}
 \includegraphics[scale=0.5]{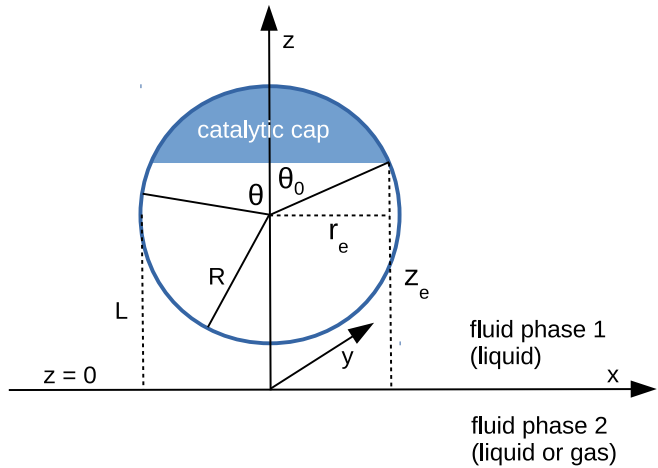}
 \caption{Schematic description of the geometry of the system. The particle, of radius 
$R$, is 
positioned with its center at a distance $L$ from the interface located at $z=0$. 
Its axis of symmetry is parallel (``cap up'', as shown in the figure) or antiparallel (``cap down'') to the normal of the interface which points into the positive $z$-direction. The size of the catalytic cap is controlled by the 
coverage angle $\theta_0$. The base of the catalytic cap forms a circular \textit{e}dge of radius $r_e$ 
at a distance $z_e$ from the interface.}
 \label{fig:0}
\end{figure}

We consider the following simple model for a chemically active 
colloid~\cite{Golestanian2005} (see Fig.~\ref{fig:0} for a schematic description of the 
system). 
The spherical colloid of radius $R$ is partially (or completely) covered by a 
catalyst which promotes the conversion of ``fuel'' molecules into product (solute) 
molecules. 
The 
particle is immersed in a solution, which is a fluid mixture composed of solvent 1 as well as fuel 
and product molecules occupying the upper half-space ($z>0$). The lower half-space ($z<0$) is occupied by fluid mixture composed of a solvent 2 as well as fuel and product molecules. We assume the two solvents to be phase separated, in thermodynamical equilibrium, and the interface between them, i.e., the plane $z = 0$, to be microscopically thin. The center of the particle is located at a distance $L > R$ from the interface.
For simplicity, we assume that both the fuel and the 
product molecules diffuse, albeit differently, in both fluids.
Furthermore, we assume that neither of the two 
species called fuel and product exhibits preferential adsorption at the interface. 
Both solutions are considered to be Newtonian fluids. The bulk viscosities of the two 
fluids are denoted as $\mu_1$ and 
$\mu_2$, respectively, and are taken to be unaffected by the densities of the fuel (which 
is kept constant) and that of the product molecules. 
The system is thought of being in contact with reservoirs of particles which 
fix the bulk number densities of each molecular species. 
The diffusion of the fuel molecules, controlled by the diffusion 
constants $D_1$ and $D_2$ in the two fluids, respectively, is considered to 
be very fast compared with the reaction rate, which corresponds to the so-called 
reaction-limited kinetics regime~\cite{Oshanin2017}. Under these assumptions, the chemical 
reaction can approximately be captured by considering the catalyst-covered part of 
the particle as an effective source of product molecules with a time- and position-independent 
rate $Q$ of solute (i.e., product) release per area of catalyst.

We restrict our study to the case in which the diffusion of the solute in the two 
fluids is sufficiently fast such that the transport of solute by advection due to 
induced hydrodynamic flow is negligible compared to the one by diffusion. 
Accordingly, the solute number density distribution is determined solely by diffusion.
Furthermore, generically in experiments with chemically active particles, the liquid 
media are aqueous solutions and the flows induced by self-phoresis correspond to very small 
Reynolds numbers. Therefore, in the following we shall describe the corresponding 
hydrodynamics by the Stokes equation for an incompressible Newtonian fluid. 

Concerning the solute we assume that the role of the interface is simply to 
provide distinct diffusion coefficients and distinct solvation energies in the two 
fluid phases. This amounts to a partitioning effect, which is associated with the 
Donnan potential (see p. $75$ in Ref.~\cite{BagotskyBook}). The transport by 
advection is negligible relative to that by diffusion (see the discussion above), 
so that at steady state the number density $c(\mathbf{r})$ of solute is the 
solution of the Laplace equation
\begin{equation}
  \nabla^2 c(\mathbf{r}) = 0,
 \label{eq:diff-eq}
\end{equation}
subject to the following boundary conditions (see the corresponding detailed discussion in 
Ref.~\cite{Dominguez2016}): \newline
(a) constant values in each half-space far from the particle: 
\begin{equation}
 \lim_{\mathbf r \rightarrow \infty}c(|\mathbf{r}|) = 
 \begin{cases}
  c_1^\infty \geq 0\,,
  & z>0\,,\\
  c_2^\infty \geq 0\,,
  & z<0\,;\\
 \end{cases}
 \label{eq:bound-cond-dens}
\end{equation}
(b) discontinuity at the interface solely due to the different solvability in the two 
media (Donnan potential):
\begin{equation}
 \lambda c(\mathbf{r})|_{z \to 0_+} = c(\mathbf{r})|_{z \to 0_-}\,,
 \label{eq:bound_cond-lambda}
\end{equation}
with 
$\lambda$ determined by
\begin{equation}
\label{eq:lambda_cinf}
\lambda  c_1^\infty = c_2^\infty\,;
\end{equation}
(c) no transport parallel to the interface, i.e., all the solute flux leaving medium 
``1'' 
enters into medium ``2'': 
\begin{equation}
 \left.\mathbf{n}_1 \cdot (- D_1 \nabla c)\right|_{z \to 0_+} = - 
 \mathbf{n}_2 \cdot \left. (-D_2  \nabla c)\right|_{z \to 0_-}\,,
 \label{eq:bound_cond-int}
\end{equation}
where $\mathbf{n}_1$ and $\mathbf{n}_2$ denote the outer normals of the domains 
${\cal D}_1 = \{z > 0\}$ (i.e., $\mathbf{n}_1=-\mathbf e_z$) and ${\cal D}_2 = \{z < 0\}$ 
(i.e., $\mathbf{n}_2=\mathbf e_z$), respectively, and $\nabla$ is the gradient 
operator;\newline
(d) at the catalyst covered part of the surface $\Sigma_p$ of the particle 
the molecular flux of solute (i.e., the number density current along the outward 
normal $\mathbf n$ of the particle surface $\Sigma_p$) equals the ``production'' rate 
of solute by the catalytic reaction, while on the rest of the 
surface it vanishes (i.e., the particle is impermeable): 
\begin{equation}
 \mathbf{n}\cdot (-D_1 \nabla c)|_{\Sigma_p} = Q f(\theta),
 \label{eq:bound_cond-flux}
\end{equation}
where $\theta$ is the polar angle (see Fig. \ref{fig:0}) and
\begin{equation}
f(\theta):=
 \begin{cases}
 1\,, &\textrm{catalyst covered part of $\Sigma_p$} \\
 0\,, &\textrm{non-catalyst part of $\Sigma_p$}\,,
 \end{cases}
 \label{eq:bound_cond-flux2}
\end{equation}
describes the coverage of the particle by catalyst.

Due to the asymmetry introduced via the partial coverage by catalyst as well as 
due to the presence of the interface, the distribution of solute around the colloidal particle 
is spatially inhomogeneous. Since the solute molecules interact with the 
colloid differently than the solvent ones, this inhomogeneity induces gradients 
in the local osmotic pressure along the surface of the particle. As a result, flow of 
the fluid and motion of the particle emerge \cite{Anderson1989,Golestanian2005}.
If the range of these molecular interactions is much smaller than the size of the 
particle, which is a plausible assumption for most of the experimental realizations, 
the effects of the 
osmotic pressure gradient are captured via a so-called phoretic slip velocity  
$\mathbf{v}_p(\mathbf{r_p})$ of the fluid relative to the surface of the particle. The 
phoretic slip is related to the local gradient of the number density of solute 
via \cite{Anderson1989,Golestanian2005}
\begin{equation}
 \mathbf{v}_p(\mathbf{r_p})=-b\nabla_\parallel c(\mathbf{r_p}) \,,
 \label{eq:def-slip-vel}
\end{equation}
where $\nabla_\parallel$ is the gradient along the surface of the particle, $b$ 
is the so-called phoretic mobility and $r_p$ denotes the points on the surface of the particle. 
The latter encodes the effective particle-solute 
interaction according to
\begin{equation}
 b =\frac{k_BT}{\mu_{1}}\int_0^\infty dh h \left( e^{-\beta\mathcal{U}(h)}-1\right)
\end{equation}
where $1/\beta=k_BT$, $k_B$ is the Boltzmann constant, $T$ the absolute temperature, 
$\mathcal{U}$ is the effective interaction potential between the particle and the solute 
relative to that between the particle and the solvent molecules, and $\mu_1$ is 
the viscosity of the fluid in which the particle is suspended. 

Once the slip velocity is provided one can set out to solve the Stokes equation under the 
proper boundary conditions at the fluid interface. Alternatively, the velocity of the 
particle in the lab reference frame can be obtained via the reciprocal 
theorem~\cite{Lorentz_original,BrennerBook,Lorentz_transl,Popescu2009}. The reciprocal 
theorem states 
that in the absence of volume forces any two 
incompressible flow fields $\mathbf{u}(\mathbf{r})$ and 
$\hat{\mathbf{u}}(\mathbf{r})$, which are 
distinct solutions of the Stokes equations \textit{within the same domain $\cal D$}, 
i.e., solutions subject to different boundary conditions but \textit{on the 
very same boundaries $\partial \cal D$},  obey the relation
\begin{equation}
\int\limits_{\partial \cal D}^{} \mathbf{u} \cdot \hat{\boldsymbol{\sigma}} \cdot 
\mathbf{n} \,dS 
= \int\limits_{\partial \cal D}^{} \hat{\mathbf{u}} \cdot \boldsymbol{\sigma} \cdot 
\mathbf{n} \,dS\,,
\label{eq:rec-theo}
\end{equation}
where $\boldsymbol{\sigma}$ and $\hat{\boldsymbol{\sigma}}$ denote the stress tensors 
corresponding to the two flow fields $\mathbf u$ and $\hat{\mathbf u}$, 
respectively, and $\mathbf{n}$ is the outward normal of $\partial \cal D$. 

However, the present system is somewhat different, in that the immiscibility condition 
separates the fluid domain into two sub-domains (on each side of the interface the 
flow velocity along the interface normal vanishes), but the flows in the two domains 
$\mathcal{D}_1$ and $\mathcal{D}_2$ are  
connected due to the requirement of continuity of the tangential stress and of 
the tangential velocity at the interface. In spite of these complications, it has been 
shown that for an infinitely large fluid domain with a planar interface the reciprocal theorem 
takes the exact same form as in the case of a particle immersed in a single fluid medium 
\cite{Sellier2011}, i.e., 
\begin{equation}
\int\limits_{\Sigma_p}^{} \mathbf{u} \cdot \hat{\boldsymbol{\sigma}} \cdot 
\mathbf{n} \,dS 
= \int\limits_{\Sigma_p}^{} \hat{\mathbf{u}} \cdot \boldsymbol{\sigma} \cdot 
\mathbf{n} \,dS\,.
\label{eq:rec-theo-final}
\end{equation}
According to Eq.~(\ref{eq:rec-theo-final}) the velocity (rotational or 
translational) of the active particle can be determined via the stress field 
in a certain ``dual'' problem typically associated with known solutions for spatially 
uniform translation or rotation.

In the following we restrict the discussion to cases in which the system 
exhibits axial symmetry, i.e., the symmetry axis of the particle coincides with the 
$z$-direction, which in turn is the normal of the interface, and we neglect the effects 
of thermal fluctuations, in particular the rotational diffusion of the axis of the active 
colloid. Accordingly, concerning the description of the present system there is a single 
unknown quantity, which is the translational velocity of the particle along the direction 
normal to the 
interface. We therefore select as the dual problem the Stokes problem of the translation of a 
chemically inert particle
with velocity $\hat{\mathbf{U}}(L)$ along the normal of the interface, due to a 
force $\hat{\mathbf{F}}$ acting on it (located at $z=L$) with a 
no-slip boundary condition at its surface, i.e.,
\begin{equation}
 \hat {\mathbf{u}}(\mathbf{r}_p) = \hat{\mathbf{U}}\,.
\end{equation}
Accounting for the phoretic slip, at the surface of the active particle one has
\begin{equation}
 \mathbf{u}(\mathbf{r}_p)=\mathbf{U}+\mathbf{v}_p(\mathbf{r_p})
\end{equation}
with $\mathbf{v}_p(\mathbf{r_p})$ defined by Eq.~(\ref{eq:def-slip-vel}) 
and $\mathbf{U}$ denoting the velocity of the active particle which points into the 
$z$-direction.
In the usual manner \cite{Sellier2011,Stone1996}, after accounting for the fact 
that the active particle is force free (i.e., $\int_{\Sigma_p} 
\mathbf{\sigma}\cdot\mathbf{n}\, dS=0$) and that $\hat{\mathbf{U}}$ and $\mathbf{U}$ do not vary 
along the surface of the particle and therefore are independent of $\mathbf{r}_p$ (even though 
they retain a dependence on the 
distance $L$ of the center of the sphere from the interface), and that $\hat{\mathbf{U}}$ 
and $\mathbf{U}$ have a $z$-component only, one arrives at
\begin{equation}
U_z \hat{F}_z  = -\int\limits_{\Sigma_p}^{} \mathbf{v}_p(\mathbf{r}_p) 
\cdot \hat{\boldsymbol{\sigma}} \cdot \mathbf{n} \,dS \\.
\label{eq:rec-theo-4}
\end{equation}

\section{Results\label{Results}}

In order to facilitate the use of Eq. (\ref{eq:rec-theo-4}), it is necessary to solve the 
diffusion equation, thus determining the distribution of solute at the surface of the 
particle, and to find the stress tensor in the auxiliary problem of a spherical particle 
moving at zero Reynolds number towards or away from a fluid-fluid interface. Both these 
problems can be solved analytically in terms of series representations in bi-polar 
coordinates (see, e.g., Refs.~\cite{Leshansky1997,Dominguez2016,LeLe80,Michelin2015,Alexander2017}). These solutions 
are concisely summarized in Appendix \ref{app:bi--polar}. Since it is difficult to 
straightforwardly gain physical insight from such series representations, here we 
focus on an analysis in the far-field approximation ($R/L \ll 1$, see Fig. 
\ref{fig:0} and the description below), which provides closed-form, easy to decipher 
and interpret results, as a means to survey the phenomenology.
The predictions of this approximate analysis are compared with the exact results obtained 
from using bi-polar coordinates, which allows one to critically asses 
(both quantitatively and qualitatively) the reliability of this far-field 
approximation.

We discuss separately two cases, first the one in which the whole surface of the 
particle is chemically active, and second the case of a Janus colloid for which only a 
spherical cap is chemically active.
The first case allows one to highlight those effects which are solely due to the way in 
which the solute partitions between the two fluid phases. The second case provides 
insight into the interplay between the above mechanism and the asymmetry in the 
distribution of the reaction sites across the surface of the particle.

\subsection{\label{far--field approach} Far-field approach}

Within the far-field approximation, the number density $c(\mathbf{r})$ and the 
stress tensor  $\hat{\boldsymbol{\sigma}}(\mathbf{r})$ are expressed in terms of 
multipole (singularity) expansions by keeping only the lowest order terms as well as 
the first set of images needed to account for the boundary conditions at the fluid-fluid 
interface. In order to keep the analysis simple, in the following we shall disregard 
contributions stemming from the images needed to enforce the boundary conditions on the 
particle surface. Rather than \textit{a priori} analyzing the reliability of this 
approximation we shall \textit{a posteriori} check it via comparison against the exact 
solutions. Indeed, the comparison of this approximate analysis with the exact solution 
shows that this truncated far-field approximation is sufficient to grasp the most 
important effects of the interface on the dynamics. (The comparison, however, also 
identifies certain qualitative discrepancies emerging from the severe truncation of 
far-field expansions (as noted previously~\cite{Popescu2017}), which underlines the 
importance of cross-checking against analytical or numerical exact results.)

We start our analysis by decomposing the stress tensor of the auxiliary problem:
\begin{equation}
\label{eq:sigma-decompos}
 \hat{\boldsymbol{\sigma}}=\hat{\boldsymbol{\sigma}}_0+\hat{\boldsymbol{\sigma}}_{im},
\end{equation}
where $\hat{\boldsymbol{\sigma}}_0$ denotes the contribution of the Stokeslet (the 
lowest order singularity), i.e., the external force $\hat{\mathbf{F}}$ acting on the 
particle and introducing its translation with velocity $\hat{\mathbf{U}}$, while 
$\hat{\boldsymbol{\sigma}}_{im}$ denotes the contribution stemming from the image system 
of the Stokeslet \cite{Blake,BrennerBook} (i.e., the lowest order contribution of the 
images). 
Concerning the first contribution, with $\mathbf{\hat{U}}=\hat{U}\mathbf{e}_z$ it is known 
that \cite{Stone1996}
\begin{equation}
 \hat{\boldsymbol{\sigma}}_0 \cdot \mathbf{n}|_{\Sigma_p} 
 =-\frac{3}{2R}\mu_1 \hat{U}\mathbf{e}_z
 \label{eq:sigma-contr-0}\,.
\end{equation}
In the far field, i.e., for $R\ll L$ the second term $\hat{\boldsymbol{\sigma}}_{im}$ 
in Eq.~(\ref{eq:sigma-decompos}) can be expanded into a Taylor series about its value at the 
center of the particle
\begin{equation}
 \hat{\boldsymbol{\sigma}}_{im}(x,y,z) = \hat{\boldsymbol{\sigma}}_{im}(0,0,L)+\delta 
 \hat{\boldsymbol{\sigma}}_{im}(x,y,z)\,.
 \label{eq:exp-sigma0}
\end{equation}
Accordingly, the deviation, $\delta \hat{\boldsymbol{\sigma}}_{im}(x,y,z)$, from $\hat{\boldsymbol{\sigma}}_{im}(0,0,L)$ is subdominant, i.e., one has
\begin{equation}
 \frac{\delta\hat{\boldsymbol{\sigma}}_{im}(x,y,z)}{\hat{\boldsymbol{\sigma}}_{im}(0,0,L)} 
\sim \mathcal{O}\left(\frac{R}{L}\right)\,.
\end{equation}
The first term on the rhs of Eq.~(\ref{eq:exp-sigma0}) can be computed explicitly 
(see Appendix~\ref{deriv-eq-im}):
\begin{equation}
 R^2\hat{\boldsymbol{\sigma}}_{im}(0,0,L)\cdot\mathbf{n} = \frac{\hat{F}_z}{4\pi}
 \frac{R^2}{L^2}\frac{\alpha}{1+\alpha}(\cos\theta)\,\mathbf{e}_z\,,
 \label{eq:sigma_im}
\end{equation}
where 
\begin{equation}
\alpha=\frac{D_2}{D_1}\,.
\label{eq:def-alpha}
\end{equation}
(We remark that, via the Stokes-Einstein relation, one has $D_{1,2}~\propto 
1/\mu_{1,2}$; hence $\alpha$ also equals the reciprocal of the ratio of the viscosities, 
i.e., $\alpha=\mu_1/\mu_2$.) 
By using $\hat{F}_z=-6\pi\mu_1 R \hat{U}_z$, employing 
spherical coordinates $z=R\cos\theta+L$, $x=R\sin\theta\cos\phi$, and 
$y=R\sin\theta\sin\phi$, and plugging Eqs.~(\ref{eq:sigma-contr-0}) and (\ref{eq:sigma_im}) into the rhs of Eq.~(\ref{eq:rec-theo-4}) one arrives at
\begin{align}
&\int\limits_{\Sigma_p}^{} \mathbf{v}_p(\mathbf{r}_p) \cdot \hat{\boldsymbol{\sigma}} \cdot 
\mathbf{n} \,dS =\nonumber\\
&\frac{\hat{F}_z}{2}\int_0^\pi\sin\theta 
v_{p}(\theta) \mathbf{e}_\theta\cdot\mathbf{e}_z 
d\theta-\frac{\hat{F}_z}{2}\frac{R^2}{L^2}\frac{\alpha}{1+\alpha}\int_{0}^{\pi} 
v_{p}(\theta) \mathbf{e}_\theta\cdot\mathbf{e}_z \sin\theta\cos\theta \,d\theta\,
\label{eq:rec-theo-final2}
\end{align}
where 
\begin{equation}
  v_{p}(\theta)\mathbf{e}_\theta\cdot\mathbf{e}_z 
= b \frac{1}{R} \left[\frac{\partial}{\partial\theta} c(R,\theta,\phi)\right] \sin\theta \,.
\label{eq:slip-vel}
\end{equation}
Equation~(\ref{eq:rec-theo-final2}) shows that the contributions to the stress tensor stemming from the images (i.e., the last term in Eq.~(\ref{eq:rec-theo-final2})) are 
sub-leading corrections to the term of leading order in $R/L$ due to the 
Stokeslet contribution (i.e., the first term in Eq.~(\ref{eq:rec-theo-final2})).

\subsection{\label{point solution} Homogeneous active colloid}

For a homogeneously covered active colloid the number density distribution of the solute around the 
particle, within the far-field expansion truncated at order 
$\mathcal{O}((R/L)^3)$, is governed by the \textit{p}oint-\textit{s}ource term and 
reads\footnote{We remind that Eq.~(\ref{conc_point_source-sph-coord}) has been 
obtained by accounting only for those images needed to enforce the boundary conditions at 
the fluid interface. Additional images are needed to enforce the 
boundary condition on the surface of the colloidal particle~\cite{Ibrahim2016}. In 
order to keep the model simple we disregard the latter ones which, in principle, 
provide additional contributions to Eq.~(\ref{conc_point_source-sph-coord}). Such 
approximate results are sufficient to describe the qualitative behavior 
of active colloids close to fluid interfaces (as discussed in the main text).}:
\begin{equation}
 \label{conc_point_source-sph-coord}
c_{ps} (R,\theta,\phi) = 
\frac{QR}{D_1}
\left[1 + 
\dfrac{1-\lambda\alpha}{1+\lambda\alpha} 
\frac{R}{2L}\left(1-\frac{R}{2L}\cos\theta\right) 
\right]+ \mathcal{O}\left(\left(\frac{R}{L}\right)^3\right)\,.
\end{equation}
By plugging the above result into Eq.~(\ref{eq:def-slip-vel}), one obtains the 
$z$-component of the slip velocity (which is needed for the rhs of Eq.~(\ref{eq:rec-theo-final2})):
\begin{align}
  v_{p}(\theta)\mathbf{e}_\theta\cdot\mathbf{e}_z &=b \frac{1}{R} \left[\frac{\partial}{\partial\theta} c(R,\theta,\phi)\right] \sin\theta \nonumber\\
  &= b\frac{Q}{D_1}
\frac{1-\lambda\alpha}{1+\lambda\alpha}\frac{R^2}{4L^2}\sin^2\theta + 
\mathcal{O} \left(\left(\frac {R}{L}\right)^3\right)\,.
  \label{eq:vel1}
\end{align}
Finally, after substituting Eqs.~(\ref{eq:rec-theo-final2}) 
and (\ref{eq:vel1}) into  Eq. (\ref{eq:rec-theo-4}) and performing the integral,  
one obtains to leading order in $R/L$ the following expression for the velocity of the 
active particle\footnote{The velocity $U_z$ is the instantaneous velocity an active 
particle will attain when its center is at distance $L$ apart from the interface. Under 
the assumptions of the model (fast diffusion of solute, quasi-steady state 
instantaneously attained) the dynamics of the system is in the overdamped regime. 
Hence, the velocity of the particle does not depend on initial conditions and equals the 
one that would be observed by fixing an active colloid at position $L$ (for example 
by an optical trap) and suddenly releasing it.}:
\begin{equation}
  U_z = -V_o\frac{1}{6}\frac{1-\lambda\alpha}{1+\lambda\alpha}\frac{R^2}{L^2}+
  \mathcal{O} \left(\left(\frac {R}{L}\right)^3\right)
  \label{eq:vel-final}
\end{equation}
where
\begin{equation}
  V_o = \frac{Q}{D_1}b
  \label{eq:def-V0}
\end{equation}
has indeed the dimension of a velocity. 
\begin{figure}
\includegraphics[scale=0.65]{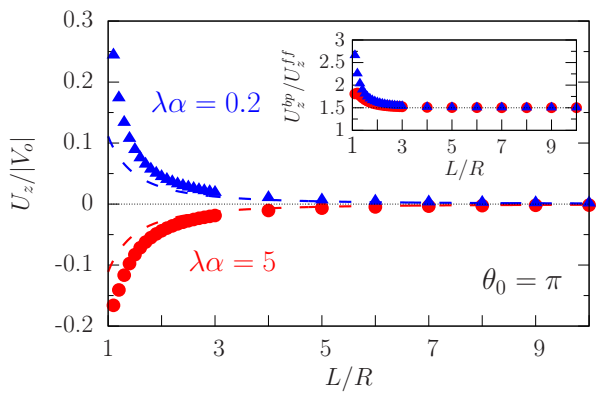}
 \caption{Rescaled velocity $U_z/|V_o|$ (see Eq.~(\ref{eq:def-V0})) as a function of the distance $L/R$ of the center of the particle from the interface. The whole surface of the particle is chemically active, i.e., $\theta_0=\pi$. The symbols (blue triangles for $\lambda\alpha=0.2$ and red circles 
 for $\lambda\alpha=5$) show the exact results calculated by using \textit{b}i-\textit{p}olar coordinates (see Appendix~\ref{app:bi--polar}), while the dashed lines show the results from the \textit{f}ar-\textit{f}ield approximation (Eq. (\ref{eq:vel-final})), for $\lambda\alpha=0.2$ 
(blue) and $\lambda\alpha=5$ (red) (see Eq.~(\ref{eq:bound_cond-lambda}) and note that $\alpha=\frac{D_2}{D_1}=\frac{\mu_1}{\mu_2}$), and $V_o<0$ (Eq.~(\ref{eq:def-V0})). The inset shows the ratio of the exact velocity $U_z^{bp}$ obtained by using \textit{b}i-\textit{p}olar coordinates and the velocity $U_z^{ff}$ calculated within the \textit{f}ar-\textit{f}ield approximation.}
 \label{fig:1}
\end{figure}

As expected from the behavior reported for similar active particles near a hard 
wall or near another particle \cite{Uspal2015,Yariv2016,Popescu2011}, in the vicinity of the 
interface the particle exhibits motion along the direction normal to the interface. Upon leaving the interface the magnitude of the 
velocity of the particle decays as $(R/L)^2$. Near the interface the velocity can reach values of the same order of 
magnitude as the maximum velocity $V_o/4$ of a Janus particle in an unbounded fluid 
\cite{Popescu2010} (see Fig. \ref{fig:1}). We note that in the limit 
$\alpha\rightarrow 0$ Eq.~(\ref{eq:vel-final}) takes the form corresponding to a hard 
wall~\cite{Uspal2015,Yariv2016}, whereas for $\alpha\rightarrow \infty$ it 
is expected that it will recover the form corresponding to a direct calculation 
for a fluid-gas interface.

Apart from the parameters included in $V_o$ (Eq.~(\ref{eq:def-V0})), the velocity 
depends on the ratio $\lambda$ of the solvabilities (Eq.~(\ref{eq:bound_cond-lambda})) and 
on the ratio $\alpha$ (Eq.~(\ref{eq:def-alpha})) of the diffusion constants of the solute 
in the two fluid phases 1 and 2 (or equivalently, due to the 
Stokes-Einstein relation, on the inverse ratio of the viscosities of the two fluids). 
These additional dependences are particularly interesting because they imply that, 
in contrast to the behavior near a hard wall, also the \textit{direction} of the motion 
depends on $\lambda \alpha$ and not solely on the sign of the phoretic mobility $b$,
which enters via $V_o$ (Eq.~(\ref{eq:def-V0})).

Taking, for example, $b<0$, which implies $V_o<0$, and  $\lambda=1$ (the discussion can 
be straightforwardly extended to the cases in which $\lambda\neq 1$ or $b>0$), 
one infers from Eq. (\ref{eq:vel-final}) that for $\alpha<1$, i.e., if the particle is 
suspended in the  \textit{less} viscous fluid phase, $U_z$ is positive and thus the 
particle moves \textit{away} from the interface. For $\alpha > 1$, i.e., if the
particle is suspended in the \textit{more} viscous fluid phase,
$U_z$ turns negative and thus the particle moves \textit{towards} the interface. 
Furthermore, Eq.~(\ref{eq:vel-final}) exhibits the symmetry relation $U_z(\lambda\alpha) = -
U_z(\frac{1}{\lambda \alpha})$. Since replacing $\lambda\alpha$ by $\frac{1}{\lambda \alpha}$ 
amounts to interchanging media 1 and 2, one concludes that, if one would perform an 
experiment in which the active particle is placed at $z = L$ and another one with an identical 
particle now placed at $z = -L$ the outcome will be the following. If in the first 
experiment the particle will move towards the interface, then in the second experiment it will 
move away from it, the speed of the motions being the same in the two cases. Vice versa, if 
in the first experiment the particle will move away from the interface, in the second if 
will move towards the interface, with the same speed of motion.

As shown in Fig.~\ref{fig:1}, the predictions of the far-field approximation, 
(Eq.~(\ref{eq:vel-final}), dashed lines), accurately capture the qualitative behavior 
of $U_z$, in particular the most important feature of the above noted change of 
sign of the velocity for $\lambda\alpha \gtrless 1$. Figure~\ref{fig:1} also shows 
that the magnitude of the velocity changes upon the replacement 
$\lambda\alpha \to 1/(\lambda\alpha)$ 
(see the asymmetry of the blue and red symbols); thus the symmetry predicted by 
Eq.~(\ref{eq:vel-final}) is an artifact of the truncation in the far-field analysis.

Quantitative discrepancies are noticeable, as highlighted by the inset of 
Fig.~\ref{fig:1}: the inset shows that, even in the 
limit $L/R\rightarrow\infty$, the amplitude of the far-field result deviates from that 
of the exact solution by ca. $50\%$. 
As it is apparent from the main panel of Figure 2, the far-field calculation correctly captures the asymptotic result of a vanishing velocity. The fact that the ratio of the far-field approximated velocity and the exact one happens in this case to be a constant, rather than a function of the distance from the interface, is somewhat peculiar. However, it merely translates into the difference between the approximated value and the exact one being 1/2 of the exact value.  Since the latter decays to zero with increasing distance from the interface, the deviation of the approximation from the exact value also vanishes with increasing distance, as it should.

We have identified two causes of these discrepancies. First, close to the interface the 
mismatch between the far-field predictions and the exact solution is due to the fact 
that we have kept only the lowest order singularities in the far-field 
approximation, while in such situations higher order terms are clearly not negligible.
Second, in the far-field approximation we have accounted 
solely for those images which are needed to enforce the 
boundary conditions at the fluid interface, while the changes induced by those 
images in the  boundary conditions at the surface of the particle have been disregarded, 
irrespectively of the distance from the interface. In this case, this kind of issue, which 
demonstrates the need for critical comparisons with exact solutions, leads to 
discrepancies between the far-field approximation and the exact result even in the 
limit $L\rightarrow \infty$ (see the inset in Fig. \ref{fig:1} and Appendix~\ref{app:dens-prof}).

\subsection{Active Janus colloid \label{Janus}}

\begin{figure}
 \includegraphics[scale=0.65]{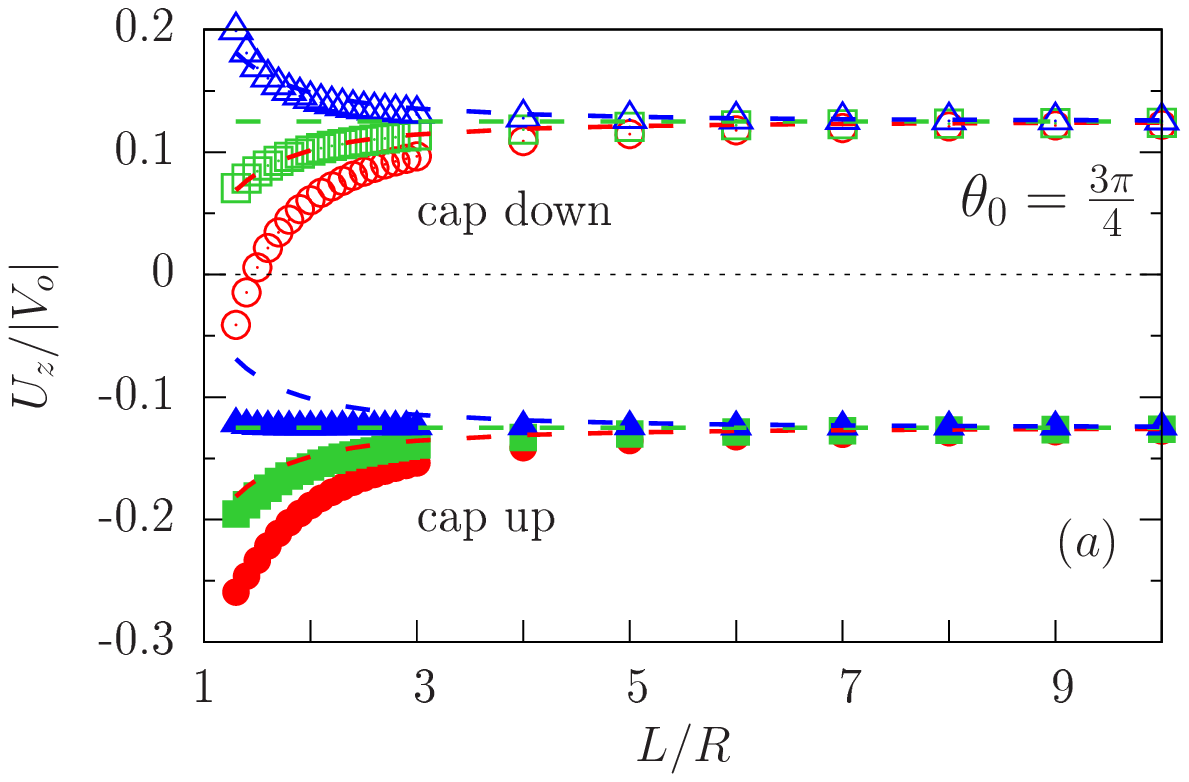}
 \includegraphics[scale=0.65]{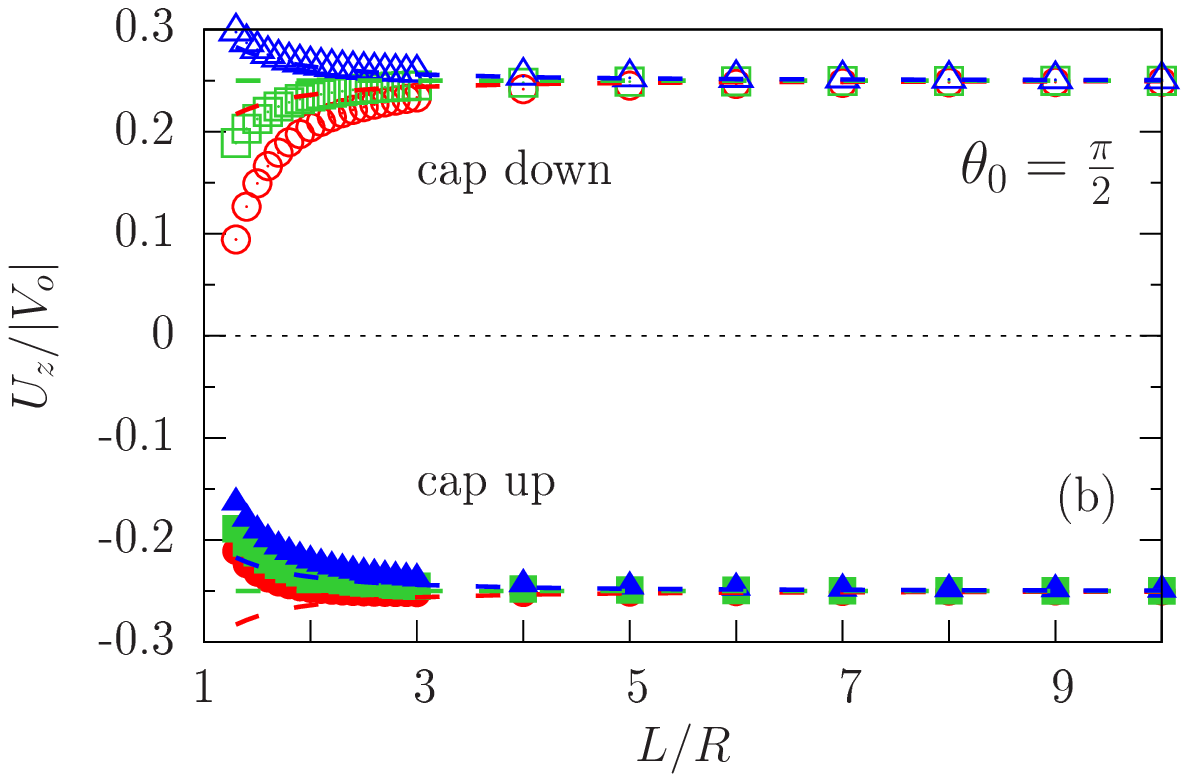}
 \includegraphics[scale=0.65]{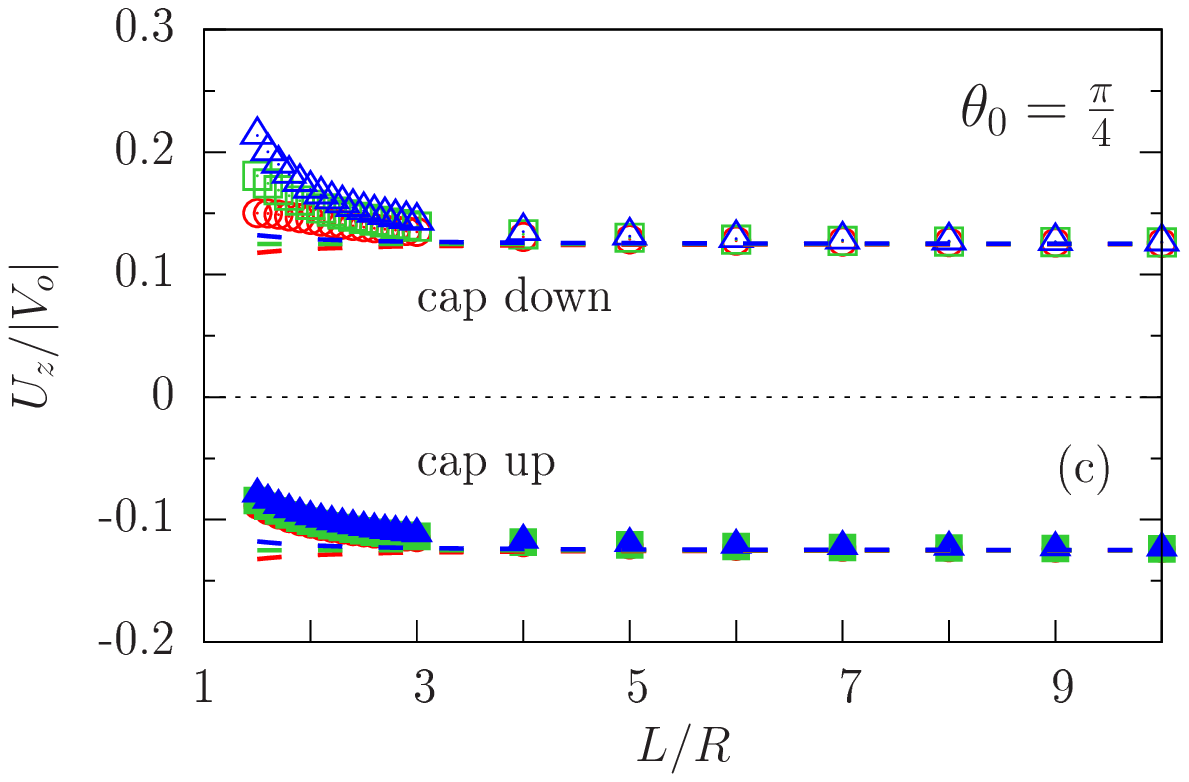}
 \caption{
Rescaled velocity $U_z/|V_o|$ calculated by using bi-polar 
coordinates (full and open symbols), as a function of the distance $L/R$ of the center of 
mass of the particle from the interface for $\theta_0=3\pi/4$ (panel (a)),  
$\theta_0=\pi/2$ (panel (b)), and $\theta_0=\pi/4$ (panel (c)) with the chemically 
active cap pointing upwards (full symbols) or downwards (open symbols) for 
$\lambda\alpha=0.2$ (blue), $\lambda\alpha=1$ (green), and $\lambda\alpha=5$ (red). 
The results shown correspond to the choice $V_o < 0$ ($b < 0$). 
The predictions of the far-field approximation are reported as dashed lines 
using the same color scheme as for the results obtained by employing bi-polar coordinates.
As expected, the comparison between Figs.~\ref{fig:1} and \ref{fig:2} shows that 
$U_z(L\rightarrow\infty,\theta_0=\pi)=0$ whereas 
$U_z(L\rightarrow\infty,\theta_0<\pi)\neq0$.
 \label{fig:2}}
\end{figure}

In this subsection we turn to the case of a Janus colloid, for which only a portion of 
the surface is chemically active. We focus on configurations in which the active side is facing either fully ''up`` 
or fully ''down``. Within the far-field approximation, the number density profile induced 
by a Janus particle comprises  a \textit{d}ipolar contribution in addition to the 
monopolar one considered in the previous subsection, i.e.,
\begin{equation}
 c(R,\theta,\phi)=c_{ps}(R,\theta,\phi)\sin^2\left(\frac{\theta_0}{2}\right)+
 c_{d}(R,\theta,\phi)\,,
\end{equation}
where the \textit{p}oint \textit{s}ource contribution $c_{ps}(R,\theta,\phi)$ 
is defined in Eq.~(\ref{conc_point_source-sph-coord}). The factor $\sin^2\left(\theta_0/2\right)$ accounts for the fact that only a portion of the surface is chemically active (Fig.~\ref{fig:0}).
Straightforward calculations (see Appendix~\ref{app-slip}) lead to the following 
expression for the dipolar contribution:\footnote{The amplitude of the image of a dipole can be 
obtained by recalling that a dipole is obtained from a pair of point sources of opposite 
magnitude located at a distance $d$ apart (see Ref.~\cite{Dominguez2016} for the 
derivation of the magnitude of the image of a point source). Accordingly, the image of a 
dipole with dipole moment $P$ located at $z=h$ is a dipole, located at $z=-h$, with dipole 
moment 
$-P\frac{1-\lambda\alpha}{1+\lambda\alpha}$ (see Appendix~\ref{app-slip}).} 
\begin{equation}
 c_{d}(R,\theta,\phi)=\frac{P}{D_1}\left[\cos\theta-
\frac{1-\lambda\alpha}{1+\lambda\alpha}\frac{R^2}{4L^2}\right]+ 
\mathcal{O}\left(\left(\frac{R}{L}\right)^3\right)\,,
\end{equation}
where (see Ref.~\cite{Uspal2016})
\begin{equation}
 P=\pm\frac{3}{8}QR\sin^2\left(\theta_0\right)\,;
 \label{eq:def-P}
\end{equation}
the plus sign ''$+$`` holds for catalytic caps pointing ''up`` while the minus sign ''$-$`` 
holds for catalytic caps pointing ''down``. 
Accordingly, the contribution of the dipole to the $z$-component of the slip 
velocity is given by
\begin{equation}
 \left(v_p(\theta)\mathbf{e}_\theta\right)\cdot\mathbf{e}_z=-\frac{P}{R D_1}b\sin^2\theta 
+ \mathcal{O}\left(\left(\frac{R}{L}\right)^3\right);
 \label{eq:dipol-final}
\end{equation}
combining this result with Eqs. (\ref{eq:def-P}) and (\ref{eq:vel-final}), one 
arrives at
\begin{equation}
 U_z=-V_o\left[\frac{1}{6}
 \frac{1-\lambda\alpha}{1+\lambda\alpha}\frac{R^2}{L^2}\sin^2\left(\frac{\theta_0}{2}
\right)\mp\frac{1}{4}\sin^2\left(\theta_0\right)\right] 
+ \mathcal{O}\left(\left(\frac{R}{L}\right)^3\right)\,
  \label{eq:vel-final-tot}
\end{equation}
where $\pm$ holds for catalytic caps ''down`` and  ''up``, respectively; note the flip of sign as compared to Eq.~(\ref{eq:def-P}). For $\theta_0=\pi$ Eq.~(\ref{eq:vel-final-tot}) reduces to Eq.~(\ref{eq:vel-final}).

The result in Eq. (\ref{eq:vel-final-tot}) deserves further discussion. Similarly 
to the symmetry discussed in the context of Eq. (\ref{eq:vel-final}), a straightforward 
calculation shows that upon simultaneously replacing $\lambda \alpha \to 1/(\lambda 
\alpha)$ and \textit{''cap up`` $\to$ ''cap down``}, $U_z$ changes sign but its magnitude 
stays the same. Thus an experiment with a particle at $z = L$ and oriented ''cap up`` 
(\textit{away from} the interface) and one in which an identical particle is immersed at 
$z = -L$ in the other fluid but oriented ''cap-down`` (\textit{towards} the interface) 
will show in one case the particle approaching the interface and in the other moving away 
from the interface, in both cases with precisely the same speed. Furthermore, it can be 
inferred that there are cases in which the two terms within the square brackets on the 
rhs of Eq. (\ref{eq:vel-final-tot}) have opposite signs. To this end we consider the 
situation of a ''cap-up`` particle, i.e., in Eq. (\ref{eq:vel-final-tot})  for the 
second term the ''-`` sign holds so that for $\lambda \alpha < 1$ the first term is positive, 
while the second is negative. In this case it is straightforward to show that if 
\begin{equation}
\label{th_crit}
\cos^2\left(\frac{\theta_0}{2}\right) \leq \cos^2\left(\frac{\theta_{cr}}{2}\right) 
 := \frac{1}{6}  \frac{1-\lambda\alpha}{1+\lambda\alpha}\,,
\end{equation}
i.e., the cap opening $\theta_0$ is larger than the critical value $\theta_{cr}$ defined 
above, then there is a particular distance $L_0$, given by 
\begin{equation}
 \label{L_stop}
 \frac{L_0}{R} = \left[\frac{\cos(\theta_{cr}/2)}{\cos(\theta_{0}/2)}\right]\,,
\end{equation}
at which the velocity of the particle is vanishing. For example, according to Eq.~(\ref{th_crit}) 
for a wall (i.e., $\alpha \to 0$) the critical opening (i.e., the size of the catalytic cap) is 
$\theta_{cr} \lesssim 3 \pi/4$. This state, provided it is stable (the stability depends also on the 
sign of $V_o$), is the equivalent of the ''hovering`` steady state for such a chemically active 
particle near a wall, as reported in Ref.~\cite{Uspal2015}. For the latter the exact critical 
value is $\theta_{cr} \simeq 0.83 \pi$ which is somewhat larger than the former far-field 
approximation. Upon increasing $\theta_0$, $\theta_{cr}$ of this state shifts towards larger 
values and, as expected, diverges in the limit $\theta_0 \to \pi$ (i.e., for a homogeneously 
active particle). The symmetry exhibited by Eq.~(\ref{eq:vel-final-tot}) and discussed above 
implies that similar states occur for a ''cap-down`` configuration for $\lambda \alpha > 1$ under 
the same constraint as stated by Eq.~(\ref{th_crit}).

Figure~\ref{fig:2} shows the dependence of $U_z/|V_o|$ for three values of 
the size of the catalytic cap and for $V_o < 0$. In particular,  Fig. \ref{fig:2}(a) 
indeed confirms that for sufficiently large coverages, such as $\theta_0 = 
3\pi/4$, active particles can be trapped at a finite distance (i.e., $U_z(L_0)=0$ for cap 
down and $\lambda \alpha = 5$), similar to what has been reported for active 
particles close to hard walls~\cite{Uspal2015}. Actually, in Fig.~\ref{fig:2}(a) the cap-down configuration 
with zero velocity is an unstable configuration in that, if the 
particle deviates from this position $L_0$, it does not return there, as can be inferred 
by inspection of the sign of the velocity to the left and to the right of the zero-crossing. 
As discussed above, such states do occur only if the coverage is sufficiently large, and 
therefore they are not observed in Figs.~\ref{fig:2}(b) and \ref{fig:2}(c). 
As in the previous case of a homogeneously active particle, the far-field 
approximation (dashed lines in  Fig.~\ref{fig:2}) misses to quantitatively capture the 
magnitude of the velocity if the active Janus particle is close to the interface -- even 
though for Janus particles it still captures the asymptotic values of $U_z$ at large 
distances $L$. As in the case of homogeneously covered particles, the symmetry 
properties inferred from Eq.~(\ref{eq:vel-final-tot}) are not confirmed by 
the exact solution, but at a qualitative level the predicted change in sign of the 
velocity upon the simultaneous change $\lambda \alpha \to 1/(\lambda 
\alpha)$ and \textit{''cap up``} $\to$ \textit{''cap down``} indeed holds for large 
$L/R$. Finally, we remark that in the case $\lambda\alpha=1$ there is no distortion of 
the number density profile due to the presence of the interface. Therefore in this 
case the variation of the velocity as function of $L$ (green symbols in   
Fig.~\ref{fig:2}) is of purely hydrodynamic origin. It stems from the boundary conditions 
imposed on the velocity profile by the fluid-fluid interface.

\section{\label{sum} Conclusions}

In order to capture the essence of the influence of a fluid-fluid interface on 
the self-diffusiophoresis of active particles, first we have studied the dynamics of an 
active colloid homogeneously covered with catalyst and being close to a fluid 
interface. For such a particle, in a homogeneous and unbounded fluid and in the absence of 
spontaneous symmetry breaking, no motion occurs. 
However, the interface breaks the translational symmetry in the transport coefficients of the products of the catalysis along the direction of the interface normal. This results in a velocity along the normal of the interface even for homogeneously covered particles. 

In order to characterize the dynamics of catalytic active particles close to a fluid-fluid 
interface we have developed  a truncated far-field expansion of the 
diffusion equation for the product molecules of the catalysis and of the Stokes 
equation. As well, an exact solution, in terms of bi-polar coordinates, has been  
constructed based on the results available in Ref.~\cite{Dominguez2016}. We have found 
that both the magnitude and the sign of the velocity of the particle can be 
controlled by tuning the ratio of the diffusivities and solvabilities of the catalysis 
products in the two coexisting fluid phases.
In particular, both the exact solution and the far-field approximation show that the sign 
of the velocity is controlled by the product $\lambda\alpha$ of the ratio $\alpha$ of the 
diffusion coefficients of the catalysis products in the two fluid phases, and the 
ratio $\lambda$ of the equilibrium solvabilities. Indeed, for a positive phoretic 
mobility $b>0$ an active particle with $\lambda\alpha>1$ moves towards the 
interface, whereas it moves away for $\lambda\alpha<1$; for $b<0$, the directions are 
reversed. This effect is strongest close to the interface; the interface-induced 
velocity decays algebraically ($\sim L^{-2}$) with the distance $L$ from the 
interface.

If the coverage of the particle is not homogeneous, there is a velocity 
$v_0=\pm\frac{\sin^2(\theta_0)}{4}V_o$ (Eq.~(\ref{eq:vel-final-tot})) already in a 
homogeneous fluid. Therefore, if such particles are close to a fluid-fluid interface 
the \textit{i}nterface induced velocity $v_i=-V_o \frac{1}{6} 
\frac{1-\lambda\alpha}{1+\lambda\alpha}\frac{R^2}{L^2}\sin^2\left(\frac{\theta_0}{2} 
\right)$ (Eq.~(\ref{eq:vel-final-tot})) sums up with $v_0$, leading to rich scenarios. 
For example, a particle with sufficiently large coverage $\theta_0$ 
(Eq. (\ref{th_crit})) experiences a stagnation point close to the interface (Eq. 
(\ref{L_stop})) where its velocity vanishes (Fig. \ref{fig:2}(a)). If stable, 
these states are the equivalent of the steady-state hovering near a planar wall. 
Clearly, for this kind of particles adsorption at the interface is hindered due to the 
interface-induced component of the velocity. On the contrary, for coverages below the 
critical value given by Eq. (\ref{th_crit}) such a tug-of-war scenario between the 
eigen-velocity of the particle and the interface-induced velocity does not occur and 
the sign of the velocity is always determined by the eigen-velocity $v_0$.

\appendix

\section{\label{app:bi--polar} Solutions in terms of bi-polar coordinates}

Both the diffusion process (Eqs. (\ref{eq:diff-eq}) - (\ref{eq:bound_cond-flux})) and the hydrodynamics of a no-slip sphere moving at zero Reynolds number towards a planar fluid-fluid interface exhibit axial symmetry and involve boundary conditions at a spherical and at a planar surface only. Accordingly, both problems can be solved exactly in terms of bi-polar coordinates. A detailed derivation of this solution is provided in the supplementary material of Ref.~\cite{Dominguez2016}. In order to be self-contained and for reasons of clarity, here we succinctly summarize the main steps and provide the formulae required for obtaining and using this solution.

\subsection*{\label{bipolar_coord}A.1: System of bi-polar coordinates}

The bi-polar coordinates $(\xi,\eta)$ with $-\infty<\xi<\infty$ and $0\leq \eta \leq \pi$ are defined such that the vertical coordinate $z$ and the radial distance $r$ from the $z$-axis are given by
\cite{BrennerBook,Bren61}
\begin{equation}
  \label{eq:bipolar}
  z = \varkappa \frac{\sinh\xi}{\cosh\xi - \cos\eta} ,
  \qquad
  r = \varkappa \frac{\sin\eta}{\cosh\xi - \cos\eta} ,
\end{equation}
where $\varkappa =  R \sinh \xi_0$ with $\xi_0 = \mathrm{arccosh}(L/R)$ is chosen 
such that the manifold $\xi = \xi_0$ corresponds to the spherical surface 
of radius $R$ centered at $z = L$ ({which is} the surface of the particle). The 
plane $z=0$ of the interface corresponds to $\xi = 0$. {In order to simplify the} 
notations we introduce the quantity $\omega := \cos \eta$. 

Here we focus on the case in which the catalytic cap, characterized by the 
opening angle $\theta_0$ (which is expressed in terms of the spherical 
coordinates attached to the particle), is oriented \textit{away} from 
the interface, as drawn in Fig.~\ref{fig:0}. (The opposite case, i.e., the 
cap facing the interface, follows from a minor change in the equations 
determining the coefficients in the expansion of the solute number density; 
this change will be pointed out at the corresponding step in the derivation.)  
In this case, the point $(\xi = \xi_0,\eta = 0)$ farthest from the interface 
(the ``north'' pole) belongs to the catalytic cap, while the point 
$(\xi = \xi_0,\eta = \pi)$ closest to the interface (the ``south'' pole) belongs 
to the chemically inert part. The boundary between the catalytic and the inert caps 
(the ``edge'') is a circle parallel to the plane $z = 0$. The points 
$P_e = (r_e,z_e)$ on the \textit{e}dge have the spherical coordinates 
$r_e = R \sin \theta_0$ and $z_e = L + R \cos \theta_0$ (see Fig.~\ref{fig:0}); 
since the edge is part of the sphere $\xi = \xi_0$, i.e., $r_e$ and $z_e$ 
satisfy Eq. (\ref{eq:bipolar}) with $\xi = \xi_0$, the points on the edge have 
the same coordinate $\eta_0$ given by 
\begin{equation}
\label{eq:eta0}
\eta_0 = \mathrm{arcctg}
\left(\frac{1 + \cos \theta_0 \cosh \xi_0}{\sin \theta_0 \sinh \xi_0} \right)\,,
\end{equation}
i.e., the edge is the intersection of the manifolds $\xi = \xi_0$ and 
$\eta = \eta_0$. Therefore, for the ``cap up'' setup the catalyst covered 
area corresponds to $(\xi_0, 0 \leq \eta \leq \eta_0)$, while for a ``cap down'' 
setup the catalyst covered area corresponds to $(\xi_0, \eta_0 \leq \eta \leq \pi)$.

\subsection*{\label{sec:diffusion} A.2: Solution of the diffusion problem}

Since all details of the corresponding calculations are provided by the 
openly accessible supplementary material of Ref.~\cite{Dominguez2016}, the brief outline given below 
for obtaining the solution of the diffusion problem and of the auxiliary stream function 
is considered to be sufficient.

The solution of the diffusion problem (Eqs.~(\ref{eq:diff-eq})-(\ref{eq:bound_cond-flux})) 
can be expressed in terms of Legendre polynomials
$P_n$ as \cite{Jeffery1912}
\begin{equation}
  \label{eq:cbipolar}
  c(\bx) = {\cal C} + \frac{Q R \sinh \xi_0}{D_1} 
  (\cosh\xi - \omega)^{1/2} 
\times \sum_{n=0}^{+\infty}
{  \left\lbrace 
    A_n \sinh \left[\left(n+\frac{1}{2}\right)\xi \right] + B_n 
\cosh \left[\left(n+\frac{1}{2}\right) \xi \right] \right \rbrace 
}
P_n(\omega)\,,~~\xi > 0\,,
\end{equation}
in fluid 1, with a similar expression but with different coefficients $\hat{\cal C}, 
\hat{A}_n$, and $\hat{B}_n$ in fluid 2 ($\xi < 0$). The prefactor $QR/D_1$ has 
the units of a number density, so that the coefficients $A_n, B_n, \hat{A}_n$, and 
$\hat{B}_n$ are dimensionless. (We note that the same prefactor $QR/D_1$ is used 
for both $\xi > 0$ and $\xi < 0$.) We focus on the solution in fluid 1 
($\xi > 0$) because only that one enters into the expression for the 
phoretic slip at the surface of the colloid.

Inserting these two series representations into the boundary conditions at infinity 
(Eqs. (\ref{eq:bound-cond-dens}) and (\ref{eq:lambda_cinf})) leads to
\begin{subequations}
 \begin{equation}
  {\cal C} = c_1^\infty,~~\hat{\cal C} = c_2^\infty = \lambda c_1^\infty\,;
 \end{equation}
inserting them into the boundary conditions at the interface (Eqs. 
(\ref{eq:bound_cond-lambda}) and  (\ref{eq:bound_cond-int})) in combination 
with the requirement that the density is bounded everywhere leads to
 \begin{equation}
  \hat{A}_n = \hat{B}_n = \lambda B_n,~ A_n = \lambda \alpha B_n,
 \end{equation}
 \label{eq:coef_diff}
\end{subequations}
with $\alpha = D_2/D_1$ (as defined in Eq. (\ref{eq:def-alpha}) in the main text). 
(Note that for a constant flux boundary condition on the particle surface, which is time- and 
position-independent over the catalyst part, the velocity of the particle turns out to be independent of the value of the constant $c_1^\infty$, see, c.f., 
Eqs. (\ref{V_bipolar_form1}) and (\ref{deriv_conc}).) By combining Eq.~(\ref{eq:coef_diff}) with the flux boundary condition (Eq.~(\ref{eq:bound_cond-flux})) at the surface of the particle and by projecting the lhs and the rhs of 
Eq. (\ref{eq:bound_cond-flux}) onto the Legendre polynomial $P_n(\omega)$, one 
arrives at the following set of linear equations determining the coefficients $B_n$:
\begin{eqnarray}
 \label{Bn_eqs}
 f_n & = & 
 (n+1) (B_n- B_{n+1}) \left\{
   \lambda \alpha \cosh\left[\left(n + \dfrac{3}{2} \right) \xi \right] 
   + 
   \sinh \left[ \left(n + \dfrac{3}{2} \right)\xi \right] 
 \right\}\nonumber\\
 & & \mbox{} + n (B_n- B_{n-1}) \left\{
   \lambda \alpha \cosh\left[\left(n - \dfrac{1}{2} \right) \xi \right] 
   + 
   \sinh \left[ \left(n - \dfrac{1}{2} \right)\xi \right] 
 \right\} ,
 \quad
 n \geq 0\,,
\end{eqnarray}
with the convention $B_{-1} = 0$. With $\omega_0 := \cos \eta_0$ defining, as 
discussed above, the edge between the active and the passive parts of the 
surface, the coefficients $f_n$ are given in terms of the activity function $f(\theta)$ 
(Eq. (\ref{eq:bound_cond-flux}) in the main text):\newline 
\begin{equation}
\label{eq:def_fn}
 f_n := (2n+1) \int\limits_{-1}^{1} d\omega 
 \dfrac{f(\omega) P_n(\omega)}{(\cosh\xi_0 - \omega)^{1/2}} 
 = 
 \begin{cases}
 (2n+1) \int\limits_{\omega_0}^{1} d\omega 
 \dfrac{P_n(\omega)}{(\cosh\xi_0 - \omega)^{1/2}}\,, & \textrm{cap up,}\\
 (2n+1) \int\limits_{-1}^{\omega_0} d\omega 
 \dfrac{P_n(\omega)}{(\cosh\xi_0 - \omega)^{1/2}}\,, & \textrm{cap down}\,.
 \end{cases}
\end{equation}
This infinitely large system of linear equations is solved by truncating it at a 
sufficiently large index $n=N$, followed by a numerical treatment. In practice, the 
truncation, as well as the series representation, are converging very fast as long as 
$L/R \gtrsim 1.1$. We have found that in most cases $N = 50$ is sufficient for 
providing accurate results. This procedure is analogous to the ones used in 
Refs. \cite{Dominguez2016} and \cite{Popescu2011}. 

\subsection*{\label{sec:hydro} A.3: Solution of the auxiliary hydrodynamics problem}

The auxiliary problem consists of a passive spherical particle, i.e., 
there is no chemical reaction, with a no slip boundary condition at its 
surface. The center of the particle is located at $z = L$ and moves with 
velocity $\hat{\mathbf{U}} = \be_z \hat{U}$ through fluid 1 along the 
direction normal to the flat fluid-fluid interface. 

The corresponding solution for the velocity field $\hat{\mathbf{u}}(\bx)$ of the 
incompressible Stokes equations can be expressed in terms of a stream function 
$\Psi_\mathrm{aux}(\bx) = \hat{U} R^2 \psi_\mathrm{aux}(\bx)$ 
as \cite{BrennerBook}
\begin{equation}
  \label{eq:stream}
  \hat{\mathbf{u}}(\bx=\br+z\be_z) = \frac{\hat{U} R^2}{r} \left[ 
    \frac{\br}{r} \frac{\partial\psi_\mathrm{aux}}{\partial z}
    - \be_z \frac{\partial\psi_\mathrm{aux}}{\partial r}
  \right]\,.
\end{equation}
This stream function can be represented in bi-polar coordinates 
\cite{Jeffery1912,Bren61}:
\begin{eqnarray}
\label{stream_func}
  \psi_\mathrm{aux}(\bx) &=& \frac{1}{(\cosh\xi-\omega)^{3/2}}  \sum_{n=1}^{+\infty}
   \left\lbrace
    K_n \, {\cosh \left[\left(n-\frac{1}{2}\right)\xi\right]}
    + L_n \, {\sinh \left[\left(n-\frac{1}{2}\right)\xi\right]}
  \right. 
\nonumber \\
   & &\left. +  M_n \, {\cosh \left[\left(n+\frac{3}{2}\right)\xi\right]}
    + N_n \, {\sinh \left[\left(n+\frac{3}{2}\right)\xi\right]}
  \right\rbrace 
\times  {\cal G}_{n+1}^{-1/2}(\omega) \, ,\;~\xi > 0 \,;
\end{eqnarray}
a similar expression, but with different coefficients ${\hat K}_n$, ${\hat L}_n$, 
${\hat M}_n$, and ${\hat N}_n$, holds for $\xi < 0$. In these equations 
\begin{equation}
 {\cal G}_{n}^{-1/2}(\omega) = \dfrac{P_{n-2}(\omega) - P_{n}(\omega)}{2 n - 1}
\end{equation}
denotes the Gegenbauer polynomial of order $n$ and degree $-1/2$ {\cite{Bren61}}. 
The dimensionless coefficients $K_n$, $L_n$, $M_n$, and $N_n$, as well as the 
hatted ones, depend on $\xi_0$ (but not on $\eta_0$) and are determined by the boundary 
conditions for the velocity field.
The requirement of a finite flow everywhere and the boundary conditions at infinity 
and at the interface lead to
\begin{subequations}
 \begin{equation}
 {\hat K}_n = {\hat L}_n = -{\hat M}_n =-{\hat N}_n\,, \qquad n \geq 1\,, 
 \end{equation}
\begin{equation}
{\hat K}_n = - \dfrac{1}{2} \left[\left(n-\dfrac{1}{2}\right) L_n +  
\left(n + \dfrac{3}{2}\right) N_n \right]\,
\qquad n \geq 1\,,
\end{equation}
\text{and}\\
\begin{equation}
K_n = \frac{\mu_2}{\mu_1} {\hat K}_n = -  \dfrac{\mu_2}{2\mu_1} 
\left[\left(n-\dfrac{1}{2}\right) L_n + \left(n + \dfrac{3}{2}\right) N_n \right] 
\,, 
\qquad n \geq 1\,,
\end{equation}
\end{subequations}
where $\mu_{1,2}$ denote the respective viscosities of the two fluid phases. By 
combining these relations with the no-slip and no-impenetrability conditions at 
the surface of the particle, the coefficients $L_n$ and $N_n$ are obtained as
\begin{subequations}
 \label{stream_coeff}
\begin{equation}
L_n = - \dfrac{\sqrt{2}}{4} (\sinh \xi_0)^2 \, n (n+1) 
\dfrac{\chi_n^{(1)} \beta_n^{(2)} - \chi_n^{(2)} \beta_n^{(1)}}
{\alpha_n^{(1)} \beta_n^{(2)} - \alpha_n^{(2)} \beta_n^{(1)}}
\end{equation}
\text{and}\nonumber
\begin{equation}
N_n = - \dfrac{\sqrt{2}}{4} (\sinh \xi_0)^2 \, n (n+1) 
\dfrac{\chi_n^{(2)} \alpha_n^{(1)} - \chi_n^{(1)} \alpha_n^{(2)}}
{\alpha_n^{(1)} \beta_n^{(2)} - \alpha_n^{(2)} \beta_n^{(1)}}\,,
\end{equation}
\end{subequations}
where
\begin{subequations}
\begin{align}
\chi_n^{(1)} &= \dfrac{e^{-(n-1/2)\xi_0}}{n-1/2} - \dfrac{e^{-(n+3/2)\xi_0}}{n+3/2}
\,,\\
\chi_n^{(2)} &= - e^{-(n-1/2)\xi_0} + e^{-(n+3/2)\xi_0}\,,\\
\alpha_n^{(1)} &= {\sinh \left[\left(n-\frac{1}{2}\right)\xi_0\right] 
+ \dfrac{\mu_2}{2\mu_1} \left(n-\frac{1}{2}\right) 
\left\lbrace \cosh \left[\left(n+\frac{3}{2}\right)\xi_0\right] - \cosh \left[
\left(n-\frac{1}{2}\right)\xi_0\right] \right\rbrace}
\,,\\
\alpha_n^{(2)} &= {\left(n-\frac{1}{2}\right) \left\lbrace 
\cosh \left[\left(n-\frac{1}{2}\right)\xi_0 \right] + \dfrac{\mu_2}{2\mu_1} 
\left\lbrace\left(n+\frac{3}{2}\right) 
\sinh \left[\left(n+\frac{3}{2}\right)\xi_0 \right] - \left(n-\frac{1}{2}\right) 
\sinh \left[\left(n-\frac{1}{2}\right)\xi_0 \right] \right\rbrace 
\right\rbrace}\,,\\
\beta_n^{(1)} &= {\sinh \left[\left(n+\frac{3}{2}\right)\xi_0 \right]
+ \dfrac{\mu_2}{2\mu_1} \left(n+\frac{3}{2}\right) 
\left\lbrace \cosh \left[\left(n+\frac{3}{2}\right)\xi_0\right] - \cosh \left[
\left(n-\frac{1}{2}\right)\xi_0 \right]\right\rbrace}\,, \\
\text{and}&\nonumber\\
\beta_n^{(2)} &= {\left(n+\frac{3}{2}\right) \left\lbrace 
\cosh \left[\left(n+\frac{3}{2}\right)\xi_0\right] + \dfrac{\mu_2}{2\mu_1} 
\left\lbrace\left(n+\frac{3}{2}\right) 
\sinh \left[\left(n+\frac{3}{2}\right)\xi_0\right] - \left(n-\frac{1}{2}\right)
\sinh \left[\left(n-\frac{1}{2}\right)\xi_0 \right]\right\rbrace\right\rbrace} \,. 
\end{align}
\end{subequations}
With noting that $\mu_2/\mu_1 = 1/\alpha$, the derivation of the stream function for 
the auxiliary problem is complete.

\subsection*{\label{sec:Vel_calc} A.4: Integral over the 
phoretic slip in bi-polar coordinates}

We start the calculation of the integral over the phoretic slip in Eq. 
(\ref{eq:rec-theo-4}) by noting that: (i) in terms of the bi-polar coordinates 
the normal to the surface $\Sigma_p$ of the particle is given by $\mathbf{n} = - 
\mathbf{e}_\xi$; (ii) 
the tangent plane to the surface of the particle is spanned by the unit vectors 
$\mathbf{e}_\eta$ and $\mathbf{e}_\phi$; and (iii) since the present problems 
exhibit axial symmetry, the solute number density as well as the stress 
tensor of the auxiliary problem are independent of $\phi$ (see the previous subsections).
Thus, the phoretic slip is given by $\mathbf{v}_p := - b \nabla_{||} 
c(\xi_0,\eta) = -b (h_\eta^{-1}|_{\xi_0}) \partial_\eta c(\xi_0,\eta) 
\mathbf{e}_\eta$, where $h_\eta = h_\xi = \varkappa (\cosh \xi -\omega)^{-1}$ 
denote the so-called scale (metric) factors corresponding to the $\eta$ and $\xi$ 
coordinates, respectively (see Eq.~(\ref{eq:bipolar}) and recall the abbreviation 
$\omega=\cos\eta$). With this Eq. (\ref{eq:rec-theo-4}) in the main text takes the 
form
\begin{eqnarray}
 \label{V_bipolar_form1}
U  &=& - \dfrac{2 \pi b}{\hat F}\int\limits_{0}^{\pi} 
\dfrac{\partial c(\xi_0,\eta)}{\partial \eta}  (\mathbf{e}_\eta \cdot 
\hat{\boldsymbol{\sigma}} \cdot \mathbf{e}_\xi) h_\phi(\xi_0,\eta) 
d\eta \nonumber\\
&=& - \dfrac{2 \pi b \varkappa}{\hat F} \int\limits_{-1}^{1} 
\dfrac{d\omega}{\cosh \xi_0 -\omega} 
\dfrac{\partial c(\xi_0,\eta)}{\partial \eta}  (\mathbf{e}_\eta \cdot 
\hat{\boldsymbol{\sigma}} \cdot \mathbf{e}_\xi)\,,
\end{eqnarray}
where $h_\phi = \varkappa \sin\eta (\cosh \xi -\omega)^{-1}$ denotes the scale 
factor corresponding to the $\phi$ coordinate. 

The terms on the right hand side of Eq.~(\ref{V_bipolar_form1}) are calculated as 
follows. First, $\hat F$ is determined from the stream function $\Psi_\mathrm{aux}~= 
\hat{U} R^2 \psi_\mathrm{aux}(\bx)$ as \cite{Jeffery1912,Bren61,Bart68,LeLe80}
\begin{eqnarray}
\label{Lam_def}
{\hat F} &=& - \dfrac{2 \sqrt{2} \pi \mu_1}{\varkappa} 
{\hat U} R^2 
\sum_{n=1}^\infty (K_n + L_n + M_n + N_n) =
- \dfrac{2 \sqrt{2} \pi  \mu_1}{\sinh \xi_0} R {\hat U}
\sum_{n=1}^\infty (L_n + N_n)\,,
\end{eqnarray}
with $L_n$ and $N_n$ given by Eq.~(\ref{stream_coeff}).

Second, the derivative of the number density at the surface of the particle is given by 
\begin{eqnarray}
 \label{deriv_conc}
 \dfrac{\partial c(\xi_0,\eta)}{\partial \eta} = 
 \dfrac{d c(\xi_0,\omega)}{d \omega} \dfrac{d\omega}{d\eta} &=& 
 - \dfrac{Q R \sinh \xi_0}{D_1}\sqrt{1-\omega^2}\sqrt{\cosh\xi_0-\omega}\\
 &\times&
 \left[-\dfrac{1}{2} (\cosh\xi_0-\omega)^{-1} \sum\limits_{n \geq 0} 
 W_n(\xi_0) P_n(\omega) 
 + \sum\limits_{n \geq 0} W_n(\xi_0) \dfrac{dP_n(\omega)}{d\omega}\right]\,,\nonumber
\end{eqnarray}
where $W_n(\xi_0):= A_n \sinh \left[\left(n+\frac{1}{2}\right)\xi_0 \right] + B_n 
\cosh \left[\left(n+\frac{1}{2}\right) \xi_0\right]$ (see Eq. (\ref{eq:cbipolar}));  
the coefficients $A_n$ and $B_n$ are determined from  Eqs. (\ref{Bn_eqs}) and 
(\ref{eq:def_fn}), as well as from the relation $A_n = (\lambda \alpha) B_n$. 

The contraction $\mathbf{e}_\eta\cdot \boldsymbol{\hat\sigma}\cdot\mathbf{e}_\xi$ of the 
stress tensor at the surface of the particle 
(which is immersed in fluid ``1'') is calculated as follows (see also 
Ref. \cite{Michelin2015}). By writing the hydrodynamic flow in the auxiliary 
problem as $\hat{\mathbf{u}} = {\hat u}_\xi \mathbf{e}_\xi + 
{\hat u}_\eta \mathbf{e}_\eta$ and by using the representation of the dyadic product 
$\nabla \mathbf{v}_\mathrm{aux}$ (which one needs in order to be able to calculate the stress 
tensor in the auxiliary problem) in terms of the general orthogonal curvilinear 
coordinates provided in Ref. \cite{BrennerBook} (see Appendix A-7, Eq. (A-7.7) therein), one 
arrives at (recalling  
$h_\xi = h_\eta$ and $\mathbf{e}_\eta \cdot \mathbf{e}_\xi = 0$)
\begin{eqnarray}
 \label{eta_xi_stres}
 \mathbf{e}_\eta \cdot 
\hat{\boldsymbol{\sigma}} \cdot \mathbf{e}_\xi &=& 
\mathbf{e}_\eta \cdot \left(\mu_1 \left[ 
      \nabla\hat{\mathbf{u}} + \left(\nabla\hat{\mathbf{u}}\right)^\dagger 
    \right] - \hat{p} \,\mathcal{I}\right)\cdot \mathbf{e}_\xi  
    = \mu_1 \mathbf{e}_\eta \cdot \left[ 
      \nabla\hat{\mathbf{u}} + \left(\nabla\hat{\mathbf{u}}\right)^\dagger 
    \right] \cdot \mathbf{e}_\xi 
    \nonumber\\
    &=& \mu_1\left[\dfrac{1}{h_\xi} \left(\dfrac{\partial {\hat u}_\xi}{\partial \eta} + 
    \dfrac{\partial {\hat u}_\eta}{\partial \xi} \right) - 
    \dfrac{1}{h_\xi^2} \left({\hat u}_\eta \dfrac{\partial h_\eta}{\partial \xi} + 
    {\hat u}_\xi
    \dfrac{\partial h_\xi}{\partial \eta} \right)\right]\\
    &=& \mu_1\left[\dfrac{\cosh \xi - \omega}{\varkappa} \left(-\sqrt{1-\omega^2}\, 
    \dfrac{\partial {\hat u}_\xi}{\partial \omega} + 
    \dfrac{\partial {\hat u}_\eta}{\partial \xi}\right) 
    + \dfrac{1}{\varkappa} \left({\hat u}_\xi \sqrt{1-\omega^2} + 
    {\hat u}_\eta \sinh\xi \right)\right]\,.\nonumber
\end{eqnarray}
The flow components $u_{\xi}$ and $u_{\eta}$ are obtained from the stream function 
$\Psi_\mathrm{aux}$ (derived in the previous section) as (see Ch. 4-4 in Ref. 
\cite{BrennerBook} and Eq. (\ref{eq:bipolar}))
\begin{subequations}\
\begin{equation}
u_\xi = - \dfrac{1}{r h_\eta} \dfrac{\partial \Psi_\mathrm{aux}}{\partial \eta} = 
\hat{U} R^2 \left(\dfrac{\cosh \xi - \omega}{\varkappa}\right)^2 
\dfrac{\partial \psi_\mathrm{aux}}{\partial \omega}
\end{equation}
and
\begin{equation}
u_\eta = \dfrac{1}{r h_\xi} \dfrac{\partial \Psi_\mathrm{aux}}{\partial \xi} = 
\dfrac{\hat{U} R^2}{\sqrt{1-\omega^2}} 
\left(\dfrac{\cosh \xi - \omega}{\varkappa}\right)^2 
\dfrac{\partial \psi_\mathrm{aux}}{\partial \xi}\,,
\end{equation}
\end{subequations}
which concludes the calculation.

\section{Derivation of Eq.~(\ref{eq:rec-theo-final2}) \label{deriv-eq-im}}

The pressure component of the stress tensor
\begin{equation}
 \hat{\boldsymbol{\sigma}}_{im}(x,y,z) = \hat{\boldsymbol{\sigma}}_{d}(x,y,z)-p(x,y,z) 
 \mathbf{\mathcal{I}}\,,
\end{equation} 
where $\mathbf{\mathcal{I}}$ is the identity matrix, renders a vanishing contribution 
$\mathbf{v}_p(\mathbf{r}_p) \cdot \mathcal{I} \cdot 
\mathbf{n}=0$ to the integral (Eq.~(\ref{eq:rec-theo-final2})) involving the slip velocity, 
because the latter is orthogonal to the normal of the surface. Accordingly, one needs to compute only the contribution due to the deviatoric stress tensor
\begin{equation}
 \hat{\boldsymbol{\sigma}}_{d}(x,y,z)=\eta\left[
 \begin{array}{ccc}
  2\partial_x\hat{v}_{d,x} & \partial_x\hat{v}_{d,y} +\partial_y\hat{v}_{d,x} & 
  \partial_x\hat{v}_{d,z} +\partial_z\hat{v}_{d,x}\\
  \partial_x\hat{v}_{d,y} +\partial_y\hat{v}_{d,x} & 2\partial_y\hat{v}_{d,y} & 
  \partial_y\hat{v}_{d,z} +\partial_z\hat{v}_{d,y}\\
  \partial_x\hat{v}_{d,z} +\partial_z\hat{v}_{d,x} & \partial_z\hat{v}_{d,y} +
  \partial_y\hat{v}_{d,z} & 2\partial_z\hat{v}_{d,z}
 \end{array}
 \right]\,,
\end{equation}
where each matrix element is a function of $x$, $y$, and $z$;
$\hat{\mathbf{v}}_{d}$ denotes the flow field due to the image system for a 
Stokeslet located on the $z$-axis at $z = L$ and oriented along the 
$z$-direction, i.e., normal to the fluid-fluid interface. It is given by \cite{Blake}
\begin{eqnarray}
 \hat{v}_{d,i} &=& \sum_{j=x,y,z}\frac{F_j}{8\pi\eta_1} \left[-\frac{1-\alpha}{1+\alpha}
\left(\frac{\delta_{ij}}{R}+\frac{R_iR_j}{R^3}\right)-\frac{\delta_{iz}}{R}-\frac{R_iR_z}{
 R^3} \right. \nonumber\\
 & + &
 \left.
 \sum_{l,k=x,y,z}\frac{2}{1+\alpha}L\left(\delta_{jl}\delta_{l 
k}-\delta_{jz}\delta_{kz}\right)\frac{\partial}{\partial 
R_k}\left[\frac{LR_i}{R^3}+\frac{\delta_{iz}}{R}+\frac{R_iR_z}{R^3}\right]\right]\,,~ 
i=x,y,z\,,
 \label{eq:blake}
\end{eqnarray}
where $\alpha$ is defined in Eq.~(\ref{eq:def-alpha}) 
and
\begin{equation}
  \mathbf{r}=(x,y,L+z)\,,\mathbf{R}=(x,y,2L+z)\,.
\end{equation}
For the present problem only the value of $\hat{\boldsymbol{\sigma}}_{d}$ at the center of the 
particle is needed. This allows one to exploit the symmetries of the derivatives of 
$\hat{\boldsymbol{v}}_{d}$ in order to simplify the algebra involved. In particular, 
one has that $\hat{\mathbf{v}}_d$ is even about the $x$ and $y$ axis, i.e.,  
\begin{equation}
 \hat{\mathbf{v}}_d(x,y,z)=\hat{\mathbf{v}}_d(-x,-y,z)\,,
\end{equation}
which implies that on the $z$-axis
\begin{eqnarray}
\hat{\mathbf{v}}_{d,x} &=& \hat{\mathbf{v}}_{d,y} = 0\,,\nonumber\\
\partial_x\hat{\mathbf{v}}_d &=& \partial_y\hat{\mathbf{v}}_d = 0\,.
\end{eqnarray}
Accordingly, the deviatoric stress tensor reduces to  
\begin{equation}
 \hat{\boldsymbol{\sigma}}_d(0,0,L)=\eta\left[
 \begin{array}{ccc}
  \phantom{2\partial_z}0\phantom{\hat{v}_{d,z}} & \phantom{2\partial_z}0\phantom{\hat{v}_{d,z}}  & \phantom{2\partial_z}0\phantom{\hat{v}_{d,z}}\\
  \phantom{2\partial_z}0\phantom{\hat{v}_{d,z}} & \phantom{2\partial_z}0\phantom{\hat{v}_{d,z}} & \phantom{2\partial_z}0\phantom{\hat{v}_{d,z}}\\
  \phantom{2\partial_z}0\phantom{\hat{v}_{d,z}} & \phantom{2\partial_z}0\phantom{\hat{v}_{d,z}} & 2\partial_z\hat{v}_{d,z}
 \end{array}
 \right]
\end{equation}
with $2\partial_z\hat{v}_{d,z}$ evaluated at $(x,y,z)=(0,0,L)$.
Equation~(\ref{eq:blake}) yields
\begin{equation}
 \hat{v}_{d,z}(0,0,L+z)=-\frac{\hat{F}_z}{2\pi\eta}\frac{1}{2L+z}\frac{1}{1+\alpha}\,,
\end{equation}
and hence
\begin{equation}
\partial_z\hat{v}_{d,z}(0,0,L)=\frac{\hat{F}_z}{2\pi\eta}\frac{1}{4L^2}\frac{1}{1+\alpha}
\,.
\end{equation}
With $\mathbf n$ given by
\begin{equation}
 \mathbf n=(\sin\theta\cos\phi,\sin\theta\sin\phi,\cos\theta)
\end{equation}
one arrives at
\begin{equation}
 \hat{\boldsymbol{\sigma}}_{im}(0,0,L)\cdot\mathbf n=\,\hat{\boldsymbol{\sigma}}_d(0,0,L)\cdot\mathbf n=(0,0,1)\frac{\hat{F}_z}{4\pi}\frac{1}{L^2}\frac{\alpha}{1+\alpha}\cos\theta 
 = 
\frac{\hat{F}_z}{4\pi}\frac{1}{L^2}\frac{\cos\theta}{1+\alpha}\,\mathbf{e}_z\,.
 \label{eq:def-sigma-im}
\end{equation}

\section{Density profiles}\label{app:dens-prof}

\begin{figure}[!ht]
 \includegraphics[scale=0.32]{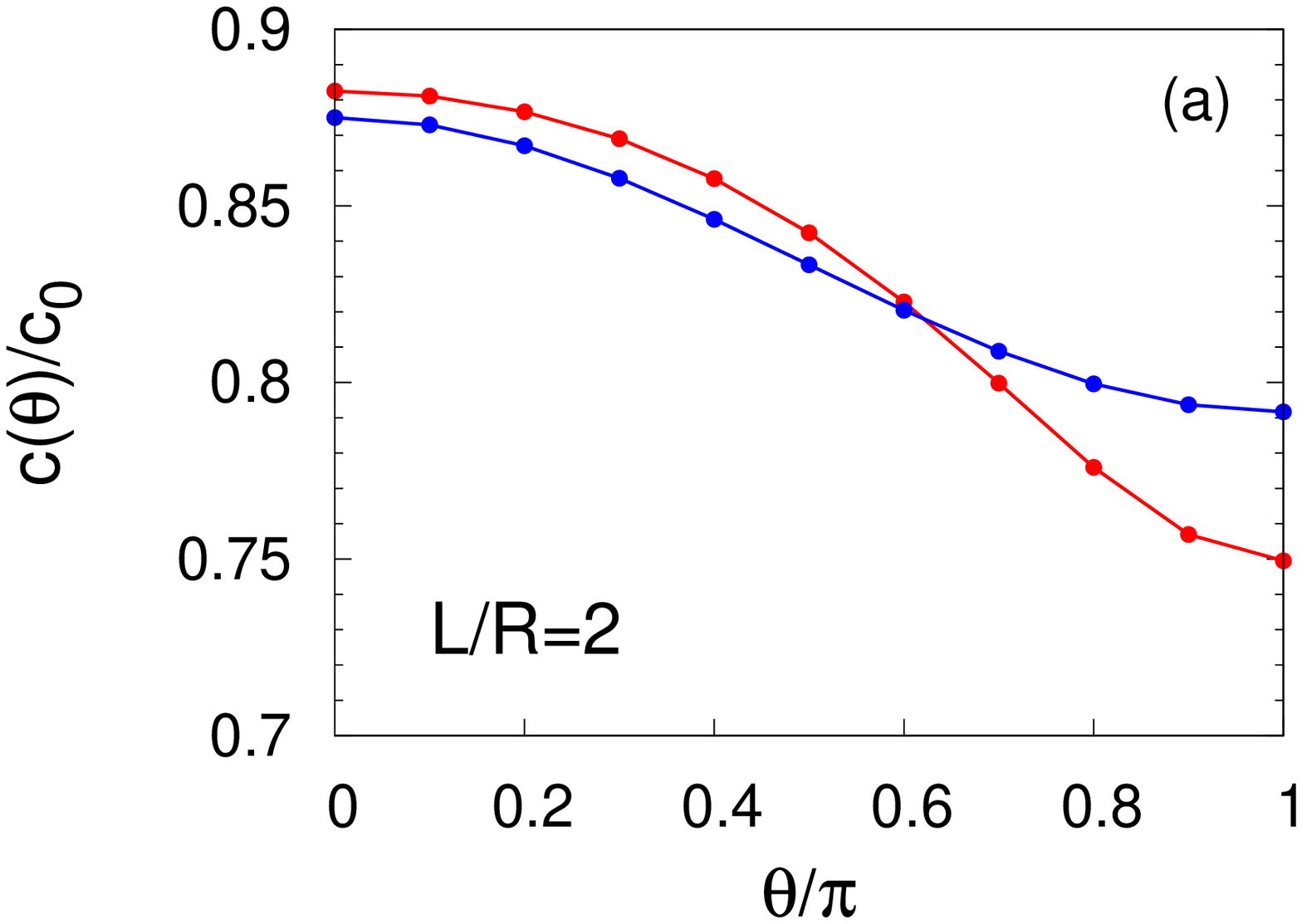}
 \includegraphics[scale=0.32]{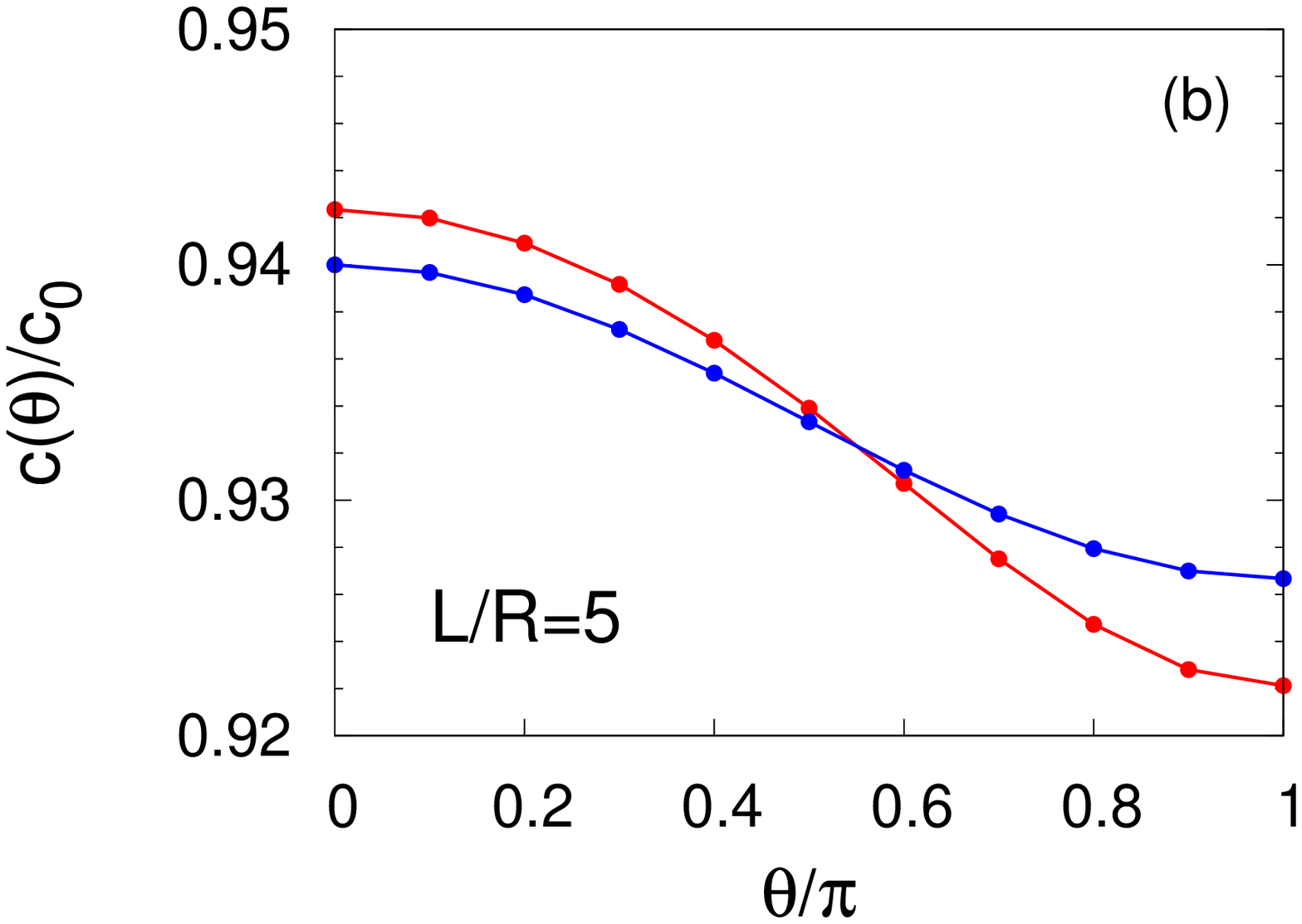}
 \includegraphics[scale=0.32]{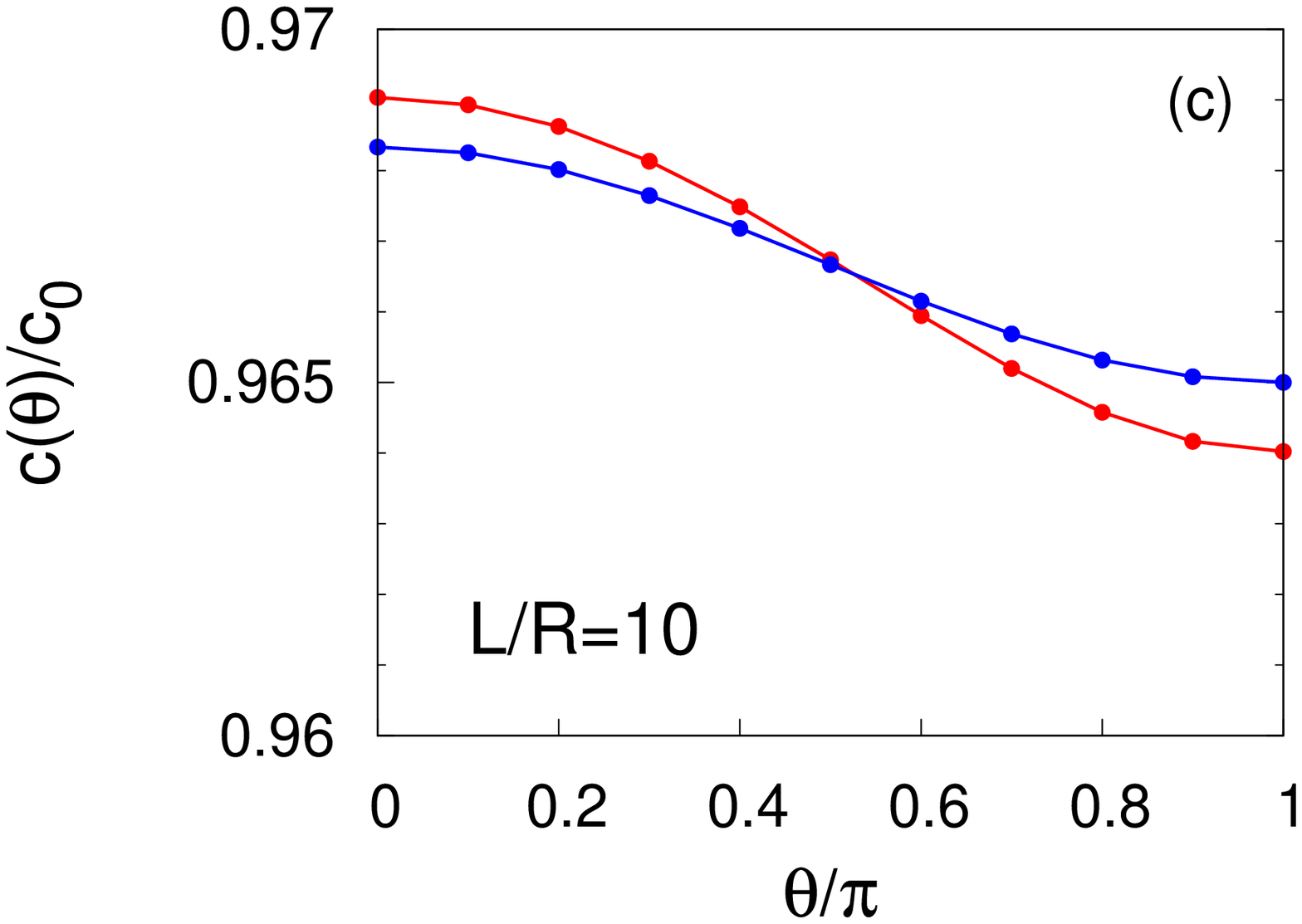}
 \caption{Number density profiles at the surface of a homogeneously covered active 
 particle near a fluid-fluid interface for $\lambda\alpha=5$ and $L/R=2$ (panel 
(a)), $L/R=5$ (panel (b)), and $L/R=10$ (panel (c)), normalized by the number 
density $c_0=QR/D_1$ at the surface of the same particle  suspended in a homogeneous 
unbounded fluid medium ($\lambda \alpha=1$). The red lines correspond to the exact 
solution (Eq.~(\ref{eq:cbipolar})) whereas the blue lines provide the far-field 
approximation (Eq.~(\ref{conc_point_source-sph-coord})).
The concentration is highest (lowest) at the north (south) pole which is 
distant (close) to the interface. The far-field approximation underestimates 
(overestimates) $c(\theta)$ on the northern (southern) hemisphere even at large 
values $L/R$ (panel (c)). Upon increasing $L/R \to \infty$ (i.e., moving the particle farther 
from the interface) both curves flatten and approach the value $c(\theta)/c_0 
\equiv 1$, as expected for a homogeneously active sphere.}
 \label{fig:app-dens}
\end{figure}

In this appendix we report the number density profiles of the products of the catalytic 
reaction for homogeneously covered active particles at various distances from the 
interface. Figure~\ref{fig:app-dens} shows the quantitative difference between the exact 
solution, as obtained by using the bi-polar coordinates, and the far-field 
approximation. 
In particular, we find that these differences persist even for large distances $L/R\gg 1$ 
from the interface. Finally, in the limit $L\rightarrow\infty$ the homogeneously 
covered particle is exposed to a homogeneous medium. In such a situation the density around the 
particle is spherically symmetric (i.e., independent of $\theta$) and the far-field expansion, 
which in this case reduces to a monopole, obviously becomes exact. Accordingly, in this situation 
the far-field predictions coincide with those of the exact solution obtained in 
bi-polar coordinates.
\newpage

\section{Slip velocity for an arbitrary multipole contribution\label{app-slip}}
\begin{figure}
\centering
 \includegraphics[scale=0.5]{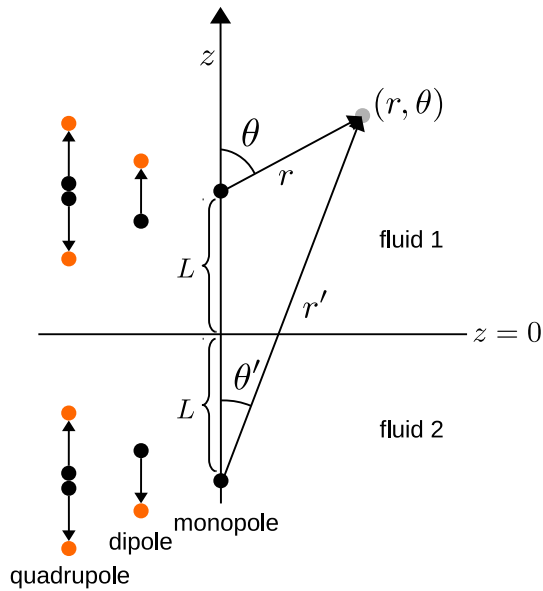}
 \caption{Schematic representation of the set of images occurring within the multipole 
 expansion. The active colloid is located at $z=L$; its first image is located at $z=-L$. 
The red dots indicate monopoles with positive amplitudes (sources) whereas 
the black dots refer to monopoles with negative amplitudes (sinks). 
Solely for visual clarity, these higher order multipoles and their images, 
located at $z = L$ and $z = -L$, respectively, are shown here in positions shifted 
to the left of the $z$-axis}
 \label{fig:ref_frame}
\end{figure}

We express the number density $c(\mathbf{r})$, with $\mathbf{r}$ in fluid $1$ and measured from the location of the particle as the sum of two series (see 
Fig.~\ref{fig:ref_frame} 
for the definition of the primed quantities):
\begin{equation}
 \label{eq:gen-sol_point_source}
c(r,\theta) = \sum_{n=0}^{\infty}c_n \left(\frac{R}{r}\right)^{n+1}P_n(\cos\theta)+
\sum_{n=0}^{\infty}c\,'_n 
\left(\frac{R}{\sqrt{r^2+4L^2+4rL\cos\theta}}\right)^{n+1}P_n(\cos\theta')
\end{equation}
where 
\begin{equation}
r'=\sqrt{r^2+4L^2+4rL\cos\theta}\,. 
\label{eq:app-r'}
\end{equation}
The coefficients $c_n$ are defined as
\begin{equation}
 c_n=\frac{2n+1}{2(n+1)}\int_0^\pi d\theta (\sin\theta) c(R,\theta)P_n(\cos\theta)
\end{equation}
and are obtained from the solution of Eqs.~(\ref{eq:diff-eq}), (\ref{eq:bound-cond-dens}), 
and (\ref{eq:bound_cond-flux}) in a homogeneous unbounded fluid, i.e., with 
$\mu_1=\mu_2$, $D_1=D_2$, $\lambda = 1$, and no interface. The coefficients $c\,'_n$ 
are the amplitudes of the images which have to be accounted for in order to fulfill the 
boundary conditions at the fluid-fluid interface, i.e., 
Eqs.~(\ref{eq:bound_cond-lambda})-(\ref{eq:bound_cond-int}).
Given the amplitude of the image for a point source, the amplitudes of the images of 
higher multipolar contributions can be derived straightforwardly by recalling that higher order 
multipoles can be obtained as sets of monopoles (see Fig.~\ref{fig:ref_frame}). 
Concerning the latter the boundary conditions in Eqs.~(\ref{eq:bound_cond-lambda}) 
and (\ref{eq:bound_cond-int}) lead to 
\begin{equation}
  c\,'_0=\frac{1-\lambda\alpha}{1+\lambda\alpha}\,.
  \label{eq:app-c0}
\end{equation}
In particular, due to the reflection symmetry of the higher order images about the 
interface (see Fig.~\ref{fig:ref_frame}), one has $c_n\times c\,'_n >0$ for $n=2i$ and $c_n\times 
c\,'_n <0$ for $n=2i+1$, $i\in\mathbb{N}_0$, respectively. Accordingly, one finds
\begin{equation}
 c\,'_n=(-1)^{n}c_n\frac{1-\lambda\alpha}{1+\lambda\alpha}\,.
\end{equation}
Substituting Eq.~(\ref{eq:gen-sol_point_source}) into Eq.~(\ref{eq:slip-vel}) renders the 
slip velocity:
\begin{eqnarray}
 v_{p}(\theta)\mathbf{e}_\theta\cdot\mathbf{e}_z&=&b\frac{1}{R}\sin\theta\left\{
 \sum_{n=1}^{\infty}c_n \frac{\partial}{\partial \theta}P_n(\cos\theta)-
 c\,'_0 \frac{2LR^2\sin\theta}{\left(R^2+4L^2+4LR\cos\theta\right)^{\frac{3}{2}}}+
 \right.\nonumber\\
 &&\left.+\sum_{n=1}^{\infty}c\,'_n 
\frac{\partial}{\partial\theta}\left[\left(\frac{R}{\sqrt{R^2+4L^2+4LR\cos\theta}}
\right)^ {n+1}P_n(\cos\theta')\right]\right\}\,.
\end{eqnarray}
For the lowest orders one finds the following:
\begin{itemize}
 \item At zeroth order, $\mathcal{O}\left(\left(\frac{R}{L}\right)^0\right)$, in the 
 expansion only the ''source`` terms, which are independent of $L$, contribute to the 
slip velocity:
 \begin{equation}
 v_{p,0}(\theta)\mathbf{e}_\theta\cdot\mathbf{e}_z=b\frac{\sin\theta}{R}\sum_{n=1}^\infty 
c_n \frac{\partial}{\partial\theta} P_n(\cos\theta)\,.
 \end{equation}
 \item At first order, $\mathcal{O}\left(\left(\frac{R}{L}\right)^1\right)$, in the 
 expansion there is no contribution to the slip velocity:
 \begin{equation}
  v_{p,1}(\theta)\mathbf{e}_\theta\cdot\mathbf{e}_z=0\,.
 \end{equation}
 \item At second order, $\mathcal{O}\left(\left(\frac{R}{L}\right)^2\right)$, in the 
 expansion the contribution to the slip velocity reads 
  \begin{equation}
  v_{p,2}(\theta)\mathbf{e}_\theta\cdot\mathbf{e}_z=\frac{b}{R}\left(\frac{R^2}{4L^2}c\,'_0-c_1\right)\sin^2\theta\,.
  \label{eq:app-v2}
 \end{equation}
 In order to obtain Eq.~(\ref{eq:app-v2}) we have used the relation 
 \begin{equation}
  \cos\theta'=\sqrt{1-\left(\frac{r}{r'}\sin\theta\right)^2}
 \end{equation}
 where $r'$ is defined in Eq.~(\ref{eq:app-r'}). Upon expanding the latter one obtains
 \begin{equation}
  \cos\theta'|_{r=R}\simeq1-\frac{R^2}{4L^2}\sin^2\theta\,.
  \label{eq:app-3}
 \end{equation}
 For a homogeneously covered particle, by using Eq.~(\ref{eq:app-c0}), Eq.~(\ref{eq:app-v2}) reduces to Eq.~(\ref{eq:vel1}) in the main text:
 \begin{equation}
  v_{p,2}(\theta)\mathbf{e}_\theta\cdot\mathbf{e}_z=b\frac{Q}{D_1}\frac{1-\lambda\alpha}{1+\lambda\alpha} \frac{R^2}{4L^2}\sin^2\theta\,
 \end{equation}
 where we have substituted $c_0=\frac{QR}{D_1}$.
 
 Alternatively, for a purely dipolar contribution Eq.~(\ref{eq:app-v2}) reduces to Eq.~(\ref{eq:dipol-final}) in the main text:
 \begin{equation}
  v_{p,2}(\theta)\mathbf{e}_\theta\cdot\mathbf{e}_z=-\frac{P}{RD_1}b\sin^2\theta\,,
 \end{equation}
 where we have substituted $c_1=\frac{P}{RD_1}$.
 \end{itemize}

\bibliography{swimmers}
 
\end{document}